# IMPROVING SIGNAL-TO-NOISE RESOLUTION IN SINGLE MOLECULE EXPERIMENTS USING MOLECULAR CONSTRUCTS WITH SHORT HANDLES

N. Forns, [†] S. de Lorenzo, [¶] M. Manosas, [‡ †] K. Hayashi, [§] J. M. Huguet, [†] and F. Ritort[†¶]*

[†]Departament de Física Fonamental, Facultat de Física, Universitat de Barcelona, 08028, Barcelona, Spain.
[‡]Laboratoire de Physique Statistique, Ecole Normale Supérieure, Unité Mixte de Recherche 8550 associée au Centre National de la Recherche Scientifique et aux Universités Paris VI et VII, 75231 Paris, France.
[§]The Institute of Scientific and Industrial Research, Osaka University, 8-1 Mihogaoka, Ibaraki 567-0047, Osaka, Japan.
[¶]CIBER de Bioingeniería, Biomateriales y Nanomedicina, Instituto de Sanidad Carlos III, Madrid, Spain.

## ABSTRACT

We investigate unfolding/folding force kinetics in DNA hairpins exhibiting two and three states with newly designed short dsDNA handles (29 bp) using optical tweezers. We show how the higher stiffness of the molecular setup moderately enhances the signal-to-noise ratio (SNR) in hopping experiments as compared to conventional long handles constructs (approximately 700 bp). The shorter construct results in a signal of higher SNR and slower folding/unfolding kinetics, thereby facilitating the detection of otherwise fast structural transitions. A novel analysis of the elastic properties of the molecular setup, based on high-bandwidth measurements of force fluctuations along the folded branch, reveals that the highest SNR that can be achieved with short handles is potentially limited by the marked reduction of the effective persistence length and stretch modulus of the short linker complex.

## INTRODUCTION

In the last years many efforts have been done to increase the resolution of different single-molecule micromanipulation techniques such as AFM, optical tweezers and magnetic tweezers (1). In all micromanipulation techniques, the molecule under study is attached to a force probe through a molecular handle. Molecular handles are used as spacers that prevent non-specific interactions between the force probe and the molecule under study. In optical tweezers experiments the molecular setup consists of the molecule of interest flanked by two handles (generally double stranded nucleic acids), one handle located at each side of the molecule, and the whole construct is tethered between two polystyrene. One bead is trapped in an optical well and is used as a force probe whereas the other bead is held fixed at the tip of a micropipette (Fig. 1 *A*). Similar constructs are used in dual trap setups (2).

Under applied force a DNA or RNA hairpin can unravel in a process reminiscent of what happens when increasing the temperature or changing the denaturant concentration. When the force applied on the hairpin is high enough (typically in the range 10-20 pN) the weak forces (hydrogen bonds plus stacking interactions) that maintain the hairpin structure are disrupted and the hairpin unfolds. Previous works with optical tweezers have investigated the folding/unfolding reaction of RNA and DNA molecules in real time by doing hopping experiments (3-5). In these experiments the molecule executes transitions between different states while the trap-pipette distance (passive mode, PM) is kept stationary or the force (constant-force mode, CFM) is maintained constant at a preset value with a feedback system. These experiments have provided accurate information about molecular folding free energies and kinetics. More recently it has been shown how handles affect the spatial and temporal resolution of single-molecule measurements (2,7,8). In the experiments dsDNA handles of 1 kb to 10 kb are typically used. An analysis of the influence of their length on the folding/unfolding kinetics (7,8) has revealed that longer handles (less stiffer) tend to give faster kinetics and lower signal-to-noise ratio (hereafter abbreviated as SNR). Current single molecule methodologies aim to use handles as stiff as possible to increase the resolution of the measurements. Although several kinds of stiff polymers might be suitable as molecular handles (e.g. carbon nanotubes (9) or microtubules (10)), the case of very short and rigid double stranded nucleic acid handles has never been studied in detail. Is it feasible to carry out single molecule experiments in the limit where handles are very short, just a few tens of bp? What are the advantages of using molecular constructs with very short handles as compared to conventional ones with long handles?

In this work we introduce a minimal construct made out of very short dsDNA handles (29 bp) used to investigate DNA folding/unfolding kinetics. Since the handles are very rigid (their contour length is five times shorter than the persistence length of dsDNA) they are expected to behave like rigid rods that fully transmit the forces to the DNA hairpin. The results here presented correspond to two DNA hairpins with different folding landscapes: (i) a hairpin that folds and unfolds in a cooperative two-states manner (2S hairpin), and (ii) a hairpin that has a fast intermediate state on-pathway (3S hairpin). We have carried out hopping experiments using a highly stable miniaturized optical tweezers (11) and we have determined the full set of kinetic parameters describing the force folding kinetics and the free energies of formation of the different structures. In general the results obtained with the new minimal construct are compatible to those obtained with the long handles (500-800 bp) conventional construct. However, the new minimal construct increases the SNR only moderately and exhibits slower folding/unfolding kinetics facilitating the detection of otherwise fast structural transitions. In order to evaluate and quantify the gain in SNR induced by the short handles we have introduced a novel method based on the analysis of high-bandwidth noise force fluctuations at different stretching forces. The method provides a way to simultaneously measure the stiffness of the optical trap and the molecular system tethered between the beads.

## MATERIALS AND METHODS

### Synthesis of DNA hairpins with short handles

The designed DNA hairpins (see Section S1 in the Supporting Material) with short handles are synthesized using the hybridization of three different oligonucleotides (Fig. 1 *B*). This method of synthesizing the short handles is easier and faster to implement as compared to the long handles synthesis. As it only requires labeling, hybridization and ligation steps one can avoid the multiple steps of synthesis of longer double stranded nucleic acids (e.g. PCR reactions, digestions with restriction enzymes, phosphorilations, dephosphorilations, DNA purifications. For the molecular setup and the synthesis of the short and long handles constructs see the procedure described in Section S2 and Fig. S2.

**Force-dependent kinetic rates**

According to Bell-Evans theory we can determine the main parameters that characterize the free energy landscape: $\Delta G_{SS'}$, $B$, $x^{\ddagger}_{SS'}$ and $x^{\ddagger}_{S'S}$ (see Section S1 and Fig. S1 for the description of the free energy landscape) by fitting the kinetic force-dependent rates to the following expressions:

(1a)
$$k_{S \to S'} = k_m e^{\beta f x^{\ddagger}_{SS'}},$$

(1b)
$$k_{S' \to S} = k_m e^{-\beta f x^{\ddagger}_{S'S} + \beta \Delta G_{SS'}},$$

where $\beta = 1/k_B T$, $k_B$ is the Boltzmann constant and $T$ the environmental temperature. S and S' stand for the folded (F), the unfolded (U) or the intermediate (I) state. $k_m$ corresponds to the unfolding rate at zero force. The term $\Delta G_{SS'}$ in Eq. 1b has been introduced to satisfy the detailed balance condition. Coexistence rates $k^C_{SS'}$ and coexistence forces $f^C_{SS'}$ are defined by $k^C_{SS'} = k_{S \to S'}(f^C_{SS'}) = k_{S' \to S}(f^C_{SS'})$. To characterize the free energy landscapes we measured the different transition rates for the DNA hairpins using PM and CFM hopping experiments (12,7,8) (see Section S3).

In all cases free energy differences, molecular extensions and coexistence forces could be also estimated from CFM and PM data by using the detailed balance property or Van't Hoff equation,

(2)
$$\Delta G_{SS'}(f) = -k_B T \log(w_{S'}/w_S) = \Delta G_{SS'} - f x_{SS'},$$

where $x_{SS'} = x^{\ddagger}_{SS'} + x^{\ddagger}_{S'S}$ and $w_S$ and $w_{S'}$ are the statistical weights of states, S and S', respectively. In PM experiments the weights in Eq. 2 are obtained from a Gaussian fit of the force distribution whereas for the CFM experiments the weights are measured from the time-dependent extension traces (see Section S4). The free energy of formation of both hairpins at zero force was obtained following the procedure described in Section S5. The procedure used to extract the mean lifetime of each state of the different hairpins from the time-dependent force traces (PM experiments) and time-dependent extension traces (CFM experiments) is described in Section S4.

**Measurement of signal-to-noise ratio**

The signal-to-noise ratio (SNR) is defined by the ratio between the jump in force in PM (or extension in CFM) in folding/unfolding transitions and the standard deviation of the signal. If $s$ denotes a generic signal (force in PM experiments or trap position in CFM) then,

(3)
$$SNR_s = \Delta s / \sigma_s,$$

where $\Delta s$ is the jump in the signal and $\sigma_s$ is the standard deviation. In order to compare the new short handles constructs with the standard long handles constructs we measured the SNR only in PM experiments (where *s* stands for the force). SNR measurements were done under PM conditions because in CFM experiments the force feedback control operates at a frequency much lower than the corner frequency of the bead. For the determination of $\sigma_f$ we collected high frequency force data at 50 kHz using a data acquisition board (National Instrument PXI-1033). The chosen bandwidth of data collection is much higher than the corner frequency of the bead at all relevant forces (around 2-3 kHz, see Section S6), a necessary condition to correctly measure force fluctuations. Measurements of $\sigma_f$ were taken for several molecules in a range of forces from 1 to 15 pN. Measurements of $\Delta f$ remained almost constant over the range of forces where hopping could be observed and was collected from hopping experiments.

## RESULTS

### Design of DNA hairpins with two and three states

We have designed a new construct (Fig. 1 *B*) that consists of a DNA hairpin inserted between two identical short dsDNA handles of 29 bp each (see Materials and Methods). These short handles are convenient to accurately follow the folding/unfolding transition as they increase the total stiffness of the system resulting in a better SNR (13,7,8). A conventional construct, having 500-800 bp handles (see Section S2) has also been synthesized to compare results obtained with both constructs. The molecular constructions are pulled applying mechanical force to the ends of the DNA molecule under study (Fig. 1 *A*).

To test the new short handles construct we investigated the force kinetics and thermodynamics of two different DNA hairpins: a two-state folder hairpin (2S) (Fig. 2 *A*); and a hairpin with an internal loop (3S hairpin) that presents an intermediate on-pathway (Fig. 3, *A*). By using the free energy values from Mfold server at 25ºC 1M NaCl (14,15,8), the free energy landscapes as a function of the molecular extension have been calculated for hairpins 2S and 3S at various forces (see Section S1). The predicted free energy landscape of the 2S hairpin (Fig. 2 *B*) at the coexistence force of 14.6 pN shows two equal free energy minima (corresponding to the folded -F- and unfolded -U- states) separated by a single free energy barrier. The F and U states are separated by $\cong 18$ nm, which is equal the extension of the released ssDNA when the hairpin unfolds at the coexistence force as measured in CFM experiments (18.2±0.9 nm, average over 8 molecules, see Fig 1 *D* and Fig. 2 *F*). The folding/unfolding reaction of the 2S hairpin can be schematically described by:

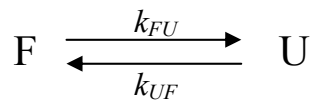

where $k_{FU}$ and $k_{UF}$ denote the force dependent unfolding and folding rates between states F and U as given by Eq. 1a and Eq. 1b (see Materials and Methods).

The predicted free energy landscape for the 3S molecule at the coexistence force between the folded and unfolded states (14.1 pN) shows the presence of an intermediate on-pathway generated by the entropy cost associated to the internal loop (Fig. 3 *B*). The sum of the distances between folded (F) and intermediate (I) and between intermediate (I) and unfolded state (U) at the coexistence force is 22.6 nm, consistent with the extension change measured during unfolding of the 3S hairpin in CFM, 22.0±1.1 nm (average over 9 molecules, see Fig. 1

*E* and Fig. 3 *F*). Four different transition rates describe the force kinetics in the 3S hairpin. These transition rates are described by the following scheme and are illustrated in Fig. 3 *B*:

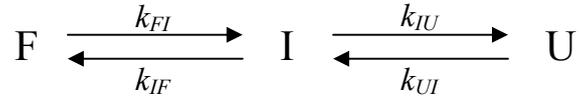

$$F \underset{k_{IF}}{\overset{k_{FI}}{\rightleftarrows}} I \underset{k_{UI}}{\overset{k_{IU}}{\rightleftarrows}} U$$

where $k_{FI}$, $k_{UI}$, $k_{IU}$ and $k_{IF}$ stand for the force dependent transition rates between states F, I and U as given by Eq. 1a and Eq. 1b (see Materials and Methods).

Pulling experiments (see Section S3), in which the force is first increased to unfold the hairpin and next decreased to allow the hairpin refolding (3,16), were initially performed with the two different constructs (short and long handles) for both hairpins. The force-distance curves (FDCs), corresponding to the measured force as a function of the distance between the center of the trap and the tip of the micropipette recorded at a pulling speed of 26 nm/s, are shown in Fig. 2 *C* and Fig. 3 *C*. The first part of the FDC corresponds to the elastic response of the dsDNA handles, and is clearly different between both constructs. The FDCs corresponding to the short handles construct show no curvature due to the high stiffness and shorter length of the handles. In contrast, the FDCs for the long handles construct show a curvature consistent with the larger elastic compliance of a longer dsDNA molecule. As the hairpin unfolds the bead in the optical trap relaxes and the tension decreases generating a force jump in the FDC. The folding/unfolding transition for the 2S molecule is a two-states transition, whereas for the 3S molecule an intermediate with a very short lifetime can be detected. The force jump obtained during the unfolding transition is nearly the same for both constructs: 1.14±0.06 pN and 1.56±0.08 pN (average over 10 molecules) for the 2S and 3S hairpins (see Fig. 2 *C* and Fig. 3 *C* insets). The force jump can be converted into molecular extension difference between the folded and unfolded conformations dividing it by the effective stiffness of the molecular setup (6). The latter is given by the slope of the FDC along the folded branch measured at the unzipping force. This gives the aforementioned 17.4±0.9 nm and 23.5±1.2 nm (average over 10 molecules) released distances for the 2S and 3S molecules respectively, the values being in agreement with the free energy landscape predictions (Fig. 2 *B* and Fig. 3 *B*). Although experiments with both handles were performed at the same pulling speed more folding/unfolding transitions along the FDC were observed with the long handles construct, suggesting that the folding/unfolding kinetics is slowed down when using shorter handles.

**Force-dependent kinetic rates measured in hopping experiments**

To study in detail the force folding/unfolding kinetics for the two hairpins we have carried out hopping experiments in the passive mode (PM) and constant force mode (CFM) (7,8,12) (see Fig. 1, *C*, *D* and *E*, Fig. 2 *D* and *F*). The unfolding and folding processes were followed by recording force traces over time (PM experiments) or extension traces over time (CFM experiments). Time traces typically span a few minutes at each condition. Since the DNA hairpin undergoes repeated cycles of folding and unfolding transitions under either mode, lifetimes in each state can be measured many times from one single experiment, making these hopping experiments useful to extract kinetic parameters such as coexistence rates $k_{SS'}^C$ and coexistence forces $f_{SS'}^C$. S and S' stand for the folded (F), the unfolded (U) or the intermediate (I) states. By applying the Bell-Evans theory we can determine the main parameters that characterize the force-dependent kinetic rates (see Materials and Methods and Section S4). In all cases free energy differences, molecular extensions and coexistence forces associated to the folding/unfolding transition could be also estimated from PM or CFM data using the detailed balance condition Eq. 2 (see Materials and Methods for details).

## I. Results for the 2S hairpin

Typical traces in the PM and CFM for the 2S hairpin are shown in Fig. 1, *C* and *D* (for short handles construct) and Fig. 2 *D* and *F* (for short and long handles constructs), and a frequency histogram for a force trace in the PM is shown in Fig. 2*E*. The hopping traces obtained with the long and short handles constructs are very similar (Fig. 2 *D* and *F*). In particular, the differences in molecular extension between the folded and the unfolded conformations extracted from the CFM experiments are equal for both constructs ($\cong 18$ nm). On the other hand, the small difference measured for the force jump in the PM (1.07±0.05 pN for long handles and 1.15±0.06 pN for short handles as obtained from averaging results over 5 molecules in both cases) is consistent with the lower effective stiffness of the long handles construct (7,8) (see below). The main difference observed between experimental traces obtained with both constructs is the presence of a higher number of folding/unfolding transitions for long handles. Indeed, at all forces measured the values obtained for kinetic rates are found to be larger with the long handles construct for both PM and CFM (see below).

The results for the fitting parameters obtained by both methods, Bell-Evans theory (Eq. 1a and Eq. 1b) (Fig. 4, *A* and *B*) and detailed balance (Eq. 2) (Fig. 4, *C* and *D*) are shown in Table 1. For the PM case, the molecular extension obtained from both methods agrees well with that measured from the PM traces and the predicted value from the free energy landscape (see Section S5). In contrast, the molecular extensions obtained from both methods in CFM experiments are larger (21-23 nm) than either the extension change directly measured from the CFM traces (~18 nm) or the predicted value (18 nm). The same effect is observed for the 3S hairpin (see next section). This artifact is consequence of the finite operational frequency of the feedback control in CFM experiments (1 kHz) (Phillip Elms, personal communication).

As expected the thermodynamic parameters are almost independent of the construct (results for short and long handles differ less than 10%), but the kinetic parameters (such as the transition rates) change with the length of the handles as previously reported (7,8). In particular, whereas the measured values of $\Delta G^0_{FU}$ in both constructs differ less than a 10% (Section S5 and Table S5), the coexistence rate measured with the long handles construct (low stiffness) is about 3-4 times higher than that measured with the short handles construct (high stiffness) (Table 1 and Fig. 4 *A* and *B*). Let us note in passing that, apart from the PM and CFM coexistence rates ($k^C_{FU}$), we can also measure the so-called apparent coexistence rate ($k^C_{app}$) in the PM experiments. These rates are the ones measured in PM but plotted as a function of the average force between the folded and unfolded states (see Section S7). It has been reported that $k^C_{app}$ decreases with the trap stiffness in DNA hopping experiments (2). This is in contrast to what happens with the coexistence rates $k^C_{FU}$ measured throughout this paper, which increase for a less stiff setup (e.g. longer handles). We have verified that whereas $k^C_{app}$ decreases with the trap stiffness, the $k^C_{FU}$ increases (see Section S7, Table S7 and Fig. S7). This is in agreement with previous studies (2).

## II. Results for the 3S hairpin

Hopping traces in the PM and CFM for 3S hairpin (Fig. 1 *E* and Fig. 3 *D* and *F*) reveal the presence of a short-lived intermediate state (mean lifetime around 10 ms). By comparing the hopping traces of long and short handles we confirm the trends observed in the 2S hairpin. The force jump between folded and intermediate state is equal to 0.91±0.05 pN (average over 6 molecules) and 0.75±0.04 pN (average over 4 molecules) for short and long handles respectively. The force jump between the intermediate and unfolded states is equal to 0.66±0.03 pN (average over 6 molecules) and 0.56±0.03 pN (average over 4 molecules) for short and long handles respectively. The shorter force jumps and the faster kinetics observed for the long handles construct are consistent with their lower effective stiffness (7,8) (see below).

By fitting the transition rates data using the Bell-Evans theory, Eq. 1a and Eq. 1b, and the detailed balance condition, Eq. 2, we can obtain the kinetic and thermodynamics parameters characterizing hairpin 3S (Fig. 5, *A* and *B* and Table 1). Most of the thermodynamic parameters obtained with short and long handles constructs are compatible (see Section S5, Table S5 and Fig.S9). Nevertheless, some differences are observed in the mean free energy differences and distances between folded, intermediate and unfolded, that lie systematically beyond two error bars. Interestingly, differences in molecular extensions between short and long handles are also observed in the values measured for extension jumps in PM traces: 23.5±1.2 nm (average over 6 molecules) and 21.0±1.1 nm (average over 4 molecules) for short and long handles respectively, showing that the 3S molecule with short handles is thermodynamically more stable than the 3S molecule with long handles. This difference is not negligible (around 15%) and might be due to irreversible fraying effects in the stem of the hairpin in the long construct that might be favored by the presence of the longer handles.
As previously observed with the 2S hairpin the transition rates are larger with softer handles: for the long handles construct $k_{FI}^C$ and $k_{IU}^C$ are about 2-3 times and 1.25 times larger than for the short handles construct. Note that the kinetics was much affected by the handles length in the case of the 2S molecule. This result is consistent with the fact that the 3S molecule presents an intermediate (see below in the Discussion and Conclusions section).

**SNR and elastic response of long and short handles**

A relevant question in our study is the understanding of how short handles constructs increase the resolution of our measurements. As explained in the introduction one expects a higher SNR for short handles. How much resolution is gained when using short handles as compared to long handles? A careful quantitative evaluation of this question is essential to assess the advantages of the new molecular construct.

**I. Signal-to-noise ratio with long and short handles**

Previous works have shown how the length of the handles, the stiffness of the optical trap and various instrumental factors influence the measured kinetics of the molecule (2,7,8). The signal-noise-ratio (SNR) of the measurements, limited by the Brownian motion of the bead $\langle \delta x^2 \rangle$, depends on the stiffness of the molecular construct attached to the bead (handles plus hairpin), $\varepsilon_x$, and that of the trap, $\varepsilon_b$, as given by the fluctuation-dissipation theorem

$$\langle \delta x^2 \rangle = k_B T / (\varepsilon_b + \varepsilon_x) \quad \text{or} \quad \langle \delta f^2 \rangle = (k_B T \varepsilon_b^2) / (\varepsilon_b + \varepsilon_x), \tag{4}$$

Note that $\varepsilon_x$ is the combined stiffness of two serially connected springs: one represented by the handles and the other by the hairpin. Along the folded branch we assume a very rigid hairpin, $\varepsilon_x$ being just the stiffness of the handle. According to Eq.(4) the softer (i.e. the longer) the handles the higher the noise. Assuming that the jump in extension or force does not depend much on the length of the handles then, according to Eqs. 3,4, the stiffer the linker the higher the SNR is. To confirm that the SNR is higher with the short handles construct, we have measured the variance of the signal from 1 pN to 15 pN (see Materials and Methods). In Fig. 6 *A* we show examples of force-time traces for the short and long handles constructs and, as shown in Fig. 6 *B*, the force variance is higher with long handles (measured for the 2S hairpin). The SNR measured for the 2S hairpin at the coexistence force is 6.2±0.3 and 8.0±0.8 (average over 3 molecules) for long and short handles, respectively.

In order to estimate the dependence of the $SNR_f$ on the length of the handles we proceed as follows. The elastic response of the handle is described by a force-extension relation of the type $f_{L_0}(x) = \hat{f}(x/L_0)$, where *x* is the molecular extension and $L_0$ is the contour length. Then $\varepsilon_h = (1/L_0) \times (\hat{f})(x/L_0)$ meaning that, at a given force *f*, the stiffness of the handle $\varepsilon_h$ is inversely proportional to the contour length. The force jump $\Delta f$ between two states at coexistence is given by $\Delta f = \varepsilon_{eff}(f_{SS'}^C) \times \Delta x$, where $\Delta x$ is the released molecular extension. The effective stiffness $\varepsilon_{eff}$ of the system is given by,

$$1/\varepsilon_{eff}(f) = (1/\varepsilon_b) + (1/\varepsilon_x(f)),$$

(5)

where $\varepsilon_b$ and $\varepsilon_x$ are the rigidities of the trap and the molecular system attached to the bead (handles plus hairpin) respectively. According to Eq. 3 (see Materials and Methods) and using Eq. 4 we find for the SNR the following expression,

$$SNR_f = \varepsilon_x \Delta x / (k_B T (\varepsilon_x + \varepsilon_b))^{1/2},$$

(6)

In the regime of coexistence forces where $\varepsilon_x \gg \varepsilon_b$ we get $SNR_f \cong (\varepsilon_x / k_B T)^{1/2} \Delta x$ showing that, in case $\varepsilon_x = \varepsilon_h \sim 1/L_0$, $SNR_f$ decreases proportionally to the square root of the contour length of the handle. Consequently the value of $SNR_f$ for short handles is expected to be 5 times larger than for long handles. Experimentally we find that the $SNR_f$ is only 1.2 times higher with the short handles construction, quite far from the expected factor of 5. Therefore there must be an additional factor that limits the total stiffness of the system $\varepsilon_x$ in the short handles case.

## II. Elastic response of long and short handles

To elucidate where the measured factor 1.2 comes from we have carried out a detailed study of the elasticity of short and long handles from our high bandwidth measurements. In addition to the force variance shown in Fig. 6 *B* we have also measured the effective stiffness $\varepsilon_{eff}(f)$ of the molecular set up, defined by $\varepsilon_{eff}(f) = \Delta f / \Delta x$, for the 2S hairpin along the folded branch in a range of forces from 1 pN up to the coexistence force $\cong$15 pN (Fig. 6 *C*). We have measured $\varepsilon_{eff}$ over that range of forces by determining the finite derivative $\Delta f/\Delta x$ along the FDC, where $\Delta f$ =1pN and $\Delta x$ is the corresponding trap displacement. The values of the effective rigidities markedly decrease for long handles, especially at low forces. By combining Eq. 4 and Eq. 5 we can solve them at each value of the force and determine the two unknown quantities $\varepsilon_b$ and $\varepsilon_x$ as a function of force, from:

$$\varepsilon_b = \varepsilon_{eff} + \left(\langle \delta f^2 \rangle / k_B T \right), \qquad (7)$$

$$\varepsilon_x = \left(\left(\varepsilon_{eff} k_B T\right) + \langle \delta f^2 \rangle\right)\left(\varepsilon_{eff} / \langle \delta f^2 \rangle\right), \qquad (8)$$

The results for $\varepsilon_b$ and $\varepsilon_x$ are shown in Fig. 6 *C*. It is known that the stiffness of an optical trap produced by a Gaussian beam exhibits non-linearities at high enough forces (17,18). In Fig. 6 *C* (middle panel) we show the results obtained for the stiffness of the trap and the results of a fit using a linear function of the force. For $\varepsilon_b$ we find: $\varepsilon_b$=0.062+0.00059$f$ (long handles) and $\varepsilon_b$=0.058+0.00066$f$ (short handles) with $\varepsilon_b$ and $f$ expressed in pN/nm and pN units respectively. Note that in the range of forces explored (1-14 pN) the stiffness of the trap shows a moderate increase from 0.06 to 0.07 pN/nm. The value of the $\varepsilon_b$ at zero force is compatible with the trap stiffness measured at high bandwidth (50 kHz) with the bead alone in the trap (no molecule attached) (see Section S6). As shown in Fig. 6 *C* (bottom panel) the stiffness of the molecular system changes from long to short handles. Close to the coexistence force (14.5 pN) the stiffness of the short construct is approximately 3 times larger than the long handles construct. Assuming that $\varepsilon_x$ gets contribution from the handles alone (the hairpin is folded) then we should expect a factor of 25 (rather than a factor of 3) between both rigidities. Where does this discrepancy come from? We have simultaneously fit the results for $\langle \delta f^2 \rangle$, $\varepsilon_{eff}$ and $\varepsilon_x$ by assuming that $\varepsilon_x$ is given alone by the elastic response of a handle described by the worm-like chain model with variable persistence length (*p*) and stretching modulus (*Y*). Our fits reveal that the persistence lengths of the handles are strongly dependent on their contour lengths: we get *p*=1.6±0.3 nm (average over 3 molecules) for the short handles and *p*=31 ± 3nm (average over 4 molecules) for the long handles. These values are markedly lower than the standard value of 45-50 nm reported for half-lambda or lambda DNA (19-24). This decrease of the persistence length has been reported in recent studies of DNA molecules a few thousand of bp (25,26) and might be consequence of the boundary conditions imposed by the fact that the ends of the tethered molecule are anchored to the beads. Indeed, the model proposed in (25) predicts persistence lengths between 20-30nm for DNA molecules of 500-800 base-pairs (close to the size of LH) attached between two beads, values that are not far from our measurement of persistence length of LH ($\cong$30 nm). For the stretching modulus we find *Y*=390 ± 40 pN (average over 4 molecules) for long handles and *Y*=16.9 ± 1.3pN (average over 3 molecules) for short handles showing a concomitant marked decrease of the enthalpic elasticity for very short DNA molecules (see Section S8 for details of the stretching modulus).

**DISCUSSION AND CONCLUSIONS**

In this work we introduced a new minimal construct for single molecule manipulation with very short (29 bp) handles. We investigated two different DNA hairpin structures using the new short handles construct and compared the results with those obtained with conventional constructs ($\approx$700 bp handles). One hairpin has been designed to behave as an ideal two-states folder (2S hairpin), whereas the other presents a fast intermediate state on-pathway (3S hairpin). In order to investigate the folding/unfolding kinetics of these hairpins we carried out hopping experiments in PM and CFM (7,8,12). As a general trend, the thermodynamic parameters (coexistence forces, molecular extensions and folding free energies) obtained from the analysis of the hopping traces for the two different constructs yields consistent results for both the 2S and 3S hairpins (see Table 1 and 2 and Fig. 4 and 5). However we found that the 3S hairpin in the long handles construct is found to be 15% shorter and less thermodynamically less stable than the short handles construct. We speculate whether the long handles induce a residual but permanent fraying at the stem of the 3S hairpin. As expected, the main differences between both constructs appear when comparing the folding/unfolding rates: the higher effective stiffness of the experimental setup in the new minimal construct leads to slower kinetics (7,8,2). Note that the noise in the force and extension traces decreases when using the short handles leading to a higher SNR (7,8,13). Therefore the usage of stiff handles might be desirable to increase the spatial or force resolution (SNR) but also the time resolution by slowing down fast structural transitions that otherwise might not be detected.

Why the coexistence rates are higher for long handles as compared to short handles? The explanation is the base-pair hopping effect discussed in (8): during the short timescale at which individual base pairs along the hairpin breath the bead in the trap does not respond and consequently the force acting on the base pair changes (it increases if the base pair forms and decreases if the base pair dissociates) slowing the overall folding-unfolding kinetics. This change in force is lower for long handles as compared to short handles making the overall kinetics faster in the former case. In addition, for the 2S hairpin the coexistence rate for the long handles construct is 3-4 times that for the short construct. For the 3S hairpin such factor is smaller: it is 2-3 times for the coexistence rate between the folded and the intermediate states and 1.25 times between the intermediate and the unfolded states. The magnitude of such factor appears to be correlated with the released molecular extension. Indeed, the larger molecular extension released by the 2S hairpin for the F-U transition ($\cong$18 nm) should be compared to the shorter extension for the F-I ($\cong$12 nm) and I-U ($\cong$10 nm) transitions in the 3S hairpin. This leads to a bigger difference in the overall folding-unfolding kinetics between short and long handles for the 2S hairpin.

If the handles are too short, might the beads interact with each other and distort the measurements? Not with the current setup. The mean excursion of a trapped bead is given by the equipartition relation, $x_{RMS} = \sqrt{k_B T/(\varepsilon_b + \varepsilon_x)}$. At 15 pN the rigidity of the DNA tether is much larger than that of the trap, $\varepsilon_x \gg \varepsilon_b$. If we take $\varepsilon_x \approx$ 1 pN/nm (see Fig 6C, lower panel) then we get $x_{RMS} \approx$ 2 nm which is 10 times smaller than the expected contour length of two times 29bp which is about 20nm. No clashing between the beads is then expected nor observed under such conditions. On the other hand, we cannot exclude the possibility that the hairpin under a tension does not interact with the beads. This effect should be important only when the length of the hairpin is larger than the length of the handles and fluctuations of the hairpin axis along the stem are big enough for the hairpin to align along the pulling direction. Our experiments show that this effect is small though.

Another remarkable result in this study concerns the elastic properties of long and short handles. Interestingly, the SNR for short handles is found to be only 1.2 times the value for long handles when we originally expected a factor of 5. How is that possible given the fact that the short handles, being 25 times shorter than long handles, are expected to be approximately 25 times more rigid? The answer lies in the measured elastic response of short and long DNA molecules tethered between two beads. The strong decrease observed in the persistence length and stretching modulus of the dsDNA handles when their contour length is reduced from $\approx$700 bp to 29 bp contributes to drastically attenuate the increase in the stiffness.

A moderate decrease in the persistence length of dsDNA molecules (a few kb long) when tethered between two beads has been already reported (25,26). However, within the present range of much shorter molecules (between 20 bp and 700 bp) these effects seem to strongly increase. According to the extensible WLC model, the compliance of a dsDNA molecule at high forces behaves like $1/\varepsilon_x = [L_0/(4p/k_BT)^{1/2}]f^{-3/2} + L_0/Y$, the first and second terms being the entropic and enthalpic contributions respectively (see Section S8). For the long handles pulled at forces ≈15 pN the sum in $1/\varepsilon_x$ is dominated by the entropic elasticity term. Whereas for the short handles the strong decrease in the stretching modulus makes the enthalpy term mostly contribute to the overall compliance. Our experimental results show that the marked decrease in persistence length and stretching modulus of the short DNA tethers strongly limits the value of the stiffness $\varepsilon_x$ of the whole molecular setup, establishing an upper bound ($\varepsilon_x \approx 1$ pN/nm) to the maximum value that we can achieve for the SNR. The observed strong decrease of the stretching modulus for the short DNA handles is suggestive of a failure of the elastic rod model applied to short DNA molecules, a result that has been recently reported from small-angle x-ray scattering measurements (27) and that should be corroborated in future mechanical experiments. However another explanation is possible (28,29). On top of the boundary effects considered in (25) at least three effects might contribute to decrease the overall stiffness of the handles (resulting into underestimated values of the persistence length and stretch modulus). First, in the process of labeling one of the handles with digoxigenins a terminal transferase reaction is used to generate a dig-labeled ssDNA flexible tail. The length of such tail will influence the effective stiffness of the tether. Second, non-specific interactions between handles and bead could also induce irreversible fraying of the handles resulting into additional single stranded ends. Finally, the biotin/streptavidin and dig/antidig bonds might contribute with an additional soft component as well. Although we do not know which effect among these is the dominant one, all them conspire to reduce the overall stiffness of the linker. This is in agreement with what we observe.

Summing up, we have introduced a new methodology of synthesizing molecular constructs with short handles that has several advantages. First, the synthesis of the new molecular construct is easier to implement as compared to the long handles synthesis. Second, this new minimal construct can be used to moderately enhance the SNR of the measurement. Third, the kinetics of short handles construct is slower, allowing us to measure fast hopping transitions that might not be detected with conventional longer constructs. Finally, we have presented a novel method to extract accurate information about the elastic properties of the molecular setup based on high bandwidth measurements of force fluctuations. This method of analysis has two main applications: it can be used to determine the stiffness of the trap and, at the same time, can be used to extract accurate information about the elastic properties of generic polymers. The current methodology could be extended to other type of handles and systems such as RNAs and proteins that exhibit more complex molecular folding landscapes.

**SUPPORTING MATERIAL**

Sections S1, S2, S3, S4, S5, S6, S7, S8, S9  Figures S1, S2, S6, S7, S7-2, S9 and Table S5, S7 and S9 are available at www.biophys.org/biophysj/supplemental/S0006-3495(XX)XXXXX-X.

Supporting Material
Document S1. SUPPORTING MATERIAL.

**Acknowledgements.** M.M. is supported by Human Frontiers Science Program

RGP0003/2007-C and EU (BioNanoSwitch). K.H. is supported by the JSPS fellowship (No.193179). J.M.H. is supported by the Spanish Research Council in Spain. F.R. is supported by grants FIS2007-3454, Icrea Academia 2008, HFSP (RGP55-2008).

**TABLE 1 Kinetic and thermodynamic parameters of the 2S and 3S hairpins**

|  | Molecular Extension | Bell-Evans | | | | | | Detailed balance | | |
|---|---|---|---|---|---|---|---|---|---|---|
|  |  | $f^C_{FU}$ | $k^C_{FU}$ | $\Delta G_{FU}$ | $x^{\ddagger}_{FU}$ | $x^{\ddagger}_{UF}$ | $x_{FU}$ | $f^C_{FU}$ | $\Delta G_{FU}$ | $x_{FU}$ |
| 2S SH PM | 17.4 ±0.9 | 14.8 ±0.7 | 1.3 ±0.2 | 63.5 ±1.3 | 9.3 ±0.5 | 8.3 ±0.4 | 17.6 ±0.9 | 14.8 ±0.7 | 63.5 ±1.4 | 17.6 ±0.9 |
| 2S SH CFM | 18.2 ±0.9 | 14.8 ±0.7 | 1.5 ±0.2 | 73.7 ±2.7 | 10.3 ±0.5 | 10.1 ±0.5 | 20.4 ±1.0 | 14.8 ±0.7 | 76.9 ±6.4 | 21.3 ±1.8 |
| 2S LH PM | 17.2 ±0.9 | 14.8 ±0.7 | 3.8 ±0.4 | 66.2 ±1.2 | 9.7 ±0.5 | 8.7 ±0.4 | 18.4 ±0.9 | 14.6 ±0.7 | 64.5 ±0.8 | 18.1 ±0.9 |
| 2S LH CFM | 18.2 ±0.9 | 14.7 ±0.7 | 4.0 ±0.4 | 82.9 ±1.3 | 12.2 ±0.6 | 10.9 ±0.6 | 23.1 ±1.2 | 14.7 ±0.7 | 82.0 ±1.0 | 22.9 ±1.2 |

|  | Molecular Extension | Bell-Evans | | | | | | | | | |
|---|---|---|---|---|---|---|---|---|---|---|---|
|  |  | $f^C_{FI}$ | $f^C_{IU}$ | $k^C_{FI}$ | $k^C_{IU}$ | $\Delta G_{FI}$ | $\Delta G_{IU}$ | $x^{\ddagger}_{FI}$ | $x^{\ddagger}_{IF}$ | $x^{\ddagger}_{IU}$ | $x^{\ddagger}_{UI}$ |
| 3S SH PM | 23.5 ±1.2 | 14.6 ±0.7 | 12.8 ±0.6 | 6.7 ±0.5 | 8.8 ±0.3 | 50.2 ±1.2 | 32.1 ±1.2 | 7.4 ±0.4 | 6.7 ±0.3 | 6.0 ±0.3 | 4.4 ±0.4 |
| 3S SH CFM | 21.6 ±1.1 | 14.3 ±0.7 | 13.0 ±0.6 | 6.7 ±0.8 | 7.6 ±1.8 | 56.0 ±5.1 | 44.9 ±3.7 | 6.4 ±0.4 | 9.7 ±1.7 | 9.3 ±0.7 | 4.9 ±0.8 |
| 3S LH PM | 21.0 ±1.1 | 14.5 ±0.7 | 12.6 ±0.6 | 18.1 ±2.1 | 11.8 ±0.6 | 46.0 ±1.0 | 29.5 ±0.5 | 8.3 ±0.4 | 4.8 ±0.3 | 5.6 ±0.3 | 4.1 ±0.2 |
| 3S LH CFM | 22.5 ±1.1 | 14.4 ±0.7 | 12.9 ±0.6 | 14.6 ±1.7 | 9.3 ±0.6 | 55.2 ±4.1 | 46.0 ±1.9 | 9.3 ±0.8 | 6.5 ±0.7 | 7.4 ±0.6 | 7.2 ±0.4 |

|  | Bell-Evans | | | Detailed balance | | | | | | |
|---|---|---|---|---|---|---|---|---|---|---|
|  | $x_{FI}$ | $x_{IU}$ | $x_{FU}$ | $f^C_{FI}$ | $f^C_{IU}$ | $\Delta G_{FI}$ | $\Delta G_{IU}$ | $x_{FI}$ | $x_{IU}$ | $x_{FU}$ |
| 3S SH PM | 14.1 ±0.7 | 10.3 ±0.5 | 24.5 ±1.2 | 14.5 ±0.7 | 12.9 ±0.6 | 52.9 ±1.4 | 30.8 ±0.9 | 15.0 ±0.8 | 9.8 ±0.5 | 24.8 ±1.2 |
| 3S SH CFM | 16.1 ±1.6 | 14.2 ±1.1 | 30.3 ±1.5 | 14.4 ±0.7 | 13.0 ±0.6 | 52.2 ±3.1 | 43.0 ±2.7 | 15.0 ±0.9 | 13.6 ±0.8 | 28.5 ±1.4 |
| 3S LH PM | 13.0 ±0.7 | 9.7 ±0.5 | 22.7 ±1.1 | 14.4 ±0.7 | 12.4 ±0.6 | 44.9 ±1.2 | 27.9 ±0.5 | 12.8 ±0.6 | 9.2 ±0.5 | 22.0 ±1.1 |
| 3S LH CFM | 15.7 ±1.2 | 14.7 ±0.7 | 30.4 ±1.5 | 14.4 ±0.7 | 12.9 ±0.6 | 54.3 ±3.4 | 43.6 ±2.9 | 15.5 ±1.0 | 13.9 ±1.0 | 29.4 ±1.6 |

The forces are given in pN, the transition rates in Hz, the energies in $k_B T$ and the molecular extensions in nm. Molecular extensions reported in first column correspond to the values directly extracted from hopping traces. The number of molecules analyzed for each molecular construction is: 5 molecules for 2S with short handles for PM experiments (2S SH PM), 2S with long handles for CFM experiments (2S LH CFM) and for 3S with long handles for PM and CFM experiments (3S LH PM and 3S LH CFM, respectively); 7 molecules for 2S with long handles for PM experiments (2S LH PM) and for 3S with short handles for PM and CFM experiments (3S SH PM and 3S SH CFM, respectively); and 4 molecules for 2S with short handles for CFM experiments (2S SH CFM). The values in light grey are calculated using Eqs. 1a and Eq. 1b and the values in dark grey are the ones calculated with Eq. 2. The numbers for each cell are the average (*top*) and the error (combination of statistical and instrumental errors) (*bottom*) of the different molecules analyzed. We have chosen as a final estimate of the error the largest between the two sources of error (statistical and instrumental).

**FIGURES**

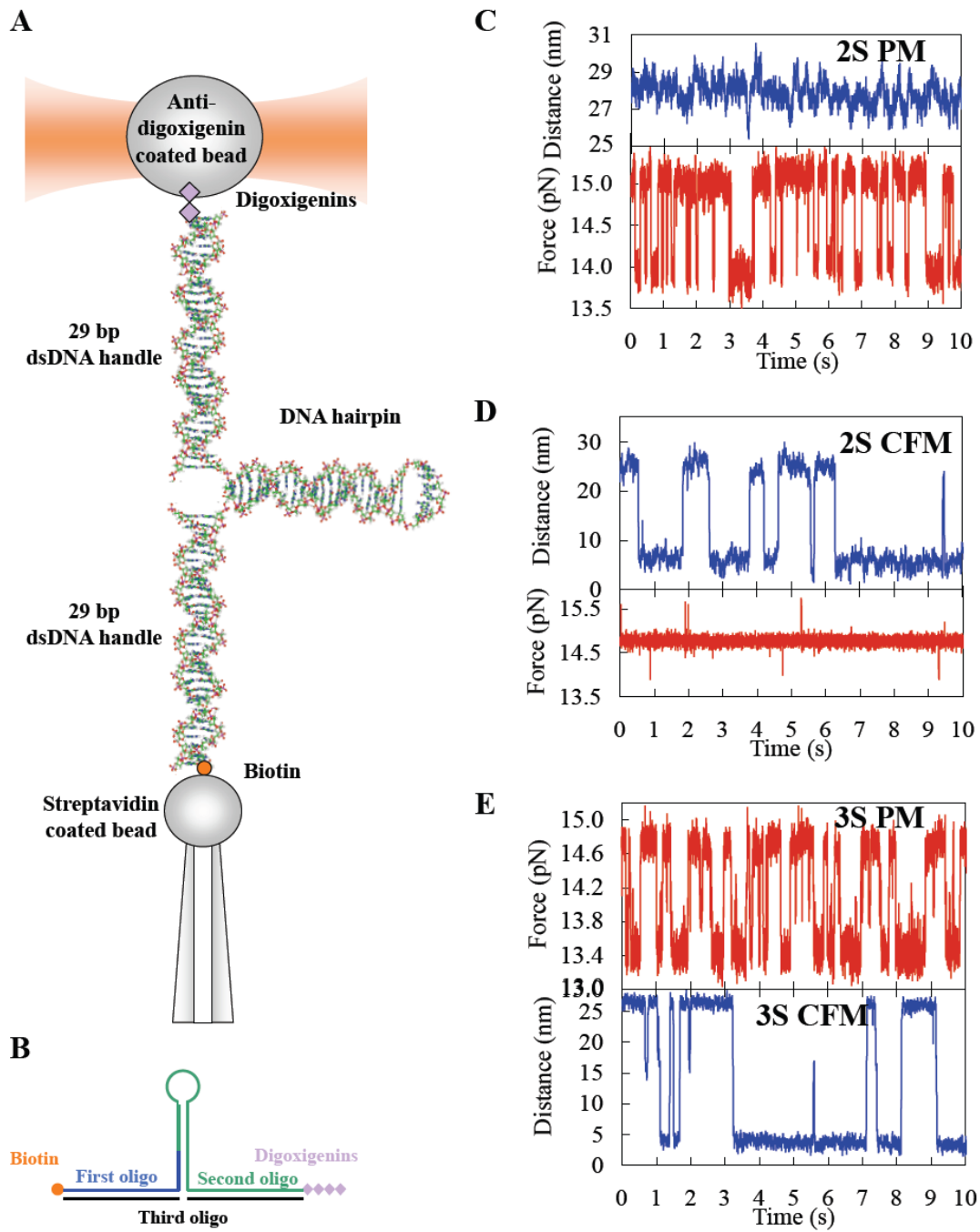

FIGURE 1 Experimental setup. (*A*) The molecular construct is attached between two beads, one bead is held by the suction of a micropipette and the other is captured in the optical trap. (*B*) Molecular construct with dsDNA short handles (29 bp each handles) made of 3 different oligonucleotides. (*C*) PM time-dependent trap-pipette relative distances (*top panel*) and time-dependent force trace (*bottom panel*) for 2S hairpin with short handles. (*D*) CFM time-dependent relative distances (*top panel*) and time-dependent force trace (*bottom panel*) for 2S hairpin with short handles. (*E*) PM time-dependent force trace (*upper panel*) and CFM time-dependent relative distances (*bottom panel*) for 3S hairpin with short handles.

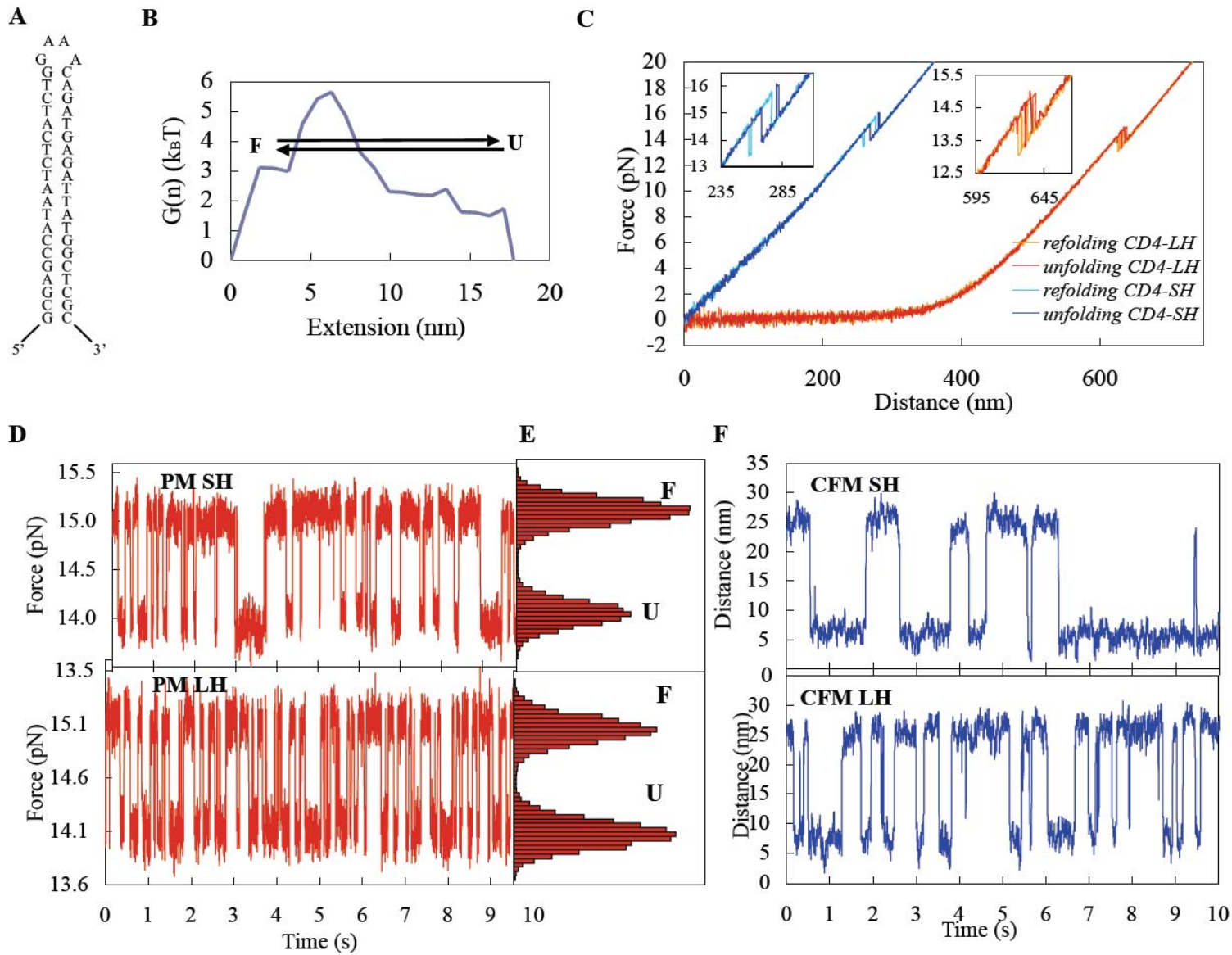

FIGURE 2 2S hairpin. (*A*) Sequence of the hairpin. (*B*) Free energy landscape plotted as a function of the molecular extension (nm) at force $f$ = 14.6 pN (at 25ºC and 1M NaCl). This was computed as described in Section S1. We also indicate the different transition rates (*arrows*). (*C*) FDC of pulling experiments with long and short handles. The insets show the unfolding and refolding of the hairpin. (*D,E*) Force traces and distributions in the PM experiments for short (*upper panels*) and long handles constructs (*lower panels*). (*F*) Time-dependent relative distances in the CFM experiments for the short (*upper panel*) and long handles constructs (*lower panel*).

FIGURE 3 3S hairpin. (*A*) Sequence of the hairpin. (*B*) Free energy landscape plotted as a function of the molecular extension (nm) at force $f = 14.1$ pN (at 25ºC and 1M NaCl). This was computed as described in Section S1. We also indicate the different transition rates (*arrows*). (*C*) FDC of pulling experiments with long and short handles. The insets show the unfolding and refolding of the hairpin. (*D,E*) Force traces and their distribution in the PM experiments for the short (*upper panels*) and long handles constructs (*lower panels*). (*F*) Time-dependent relative distances in the CFM experiments for the short (*upper panel*) and long handles constructs (*lower panel*).

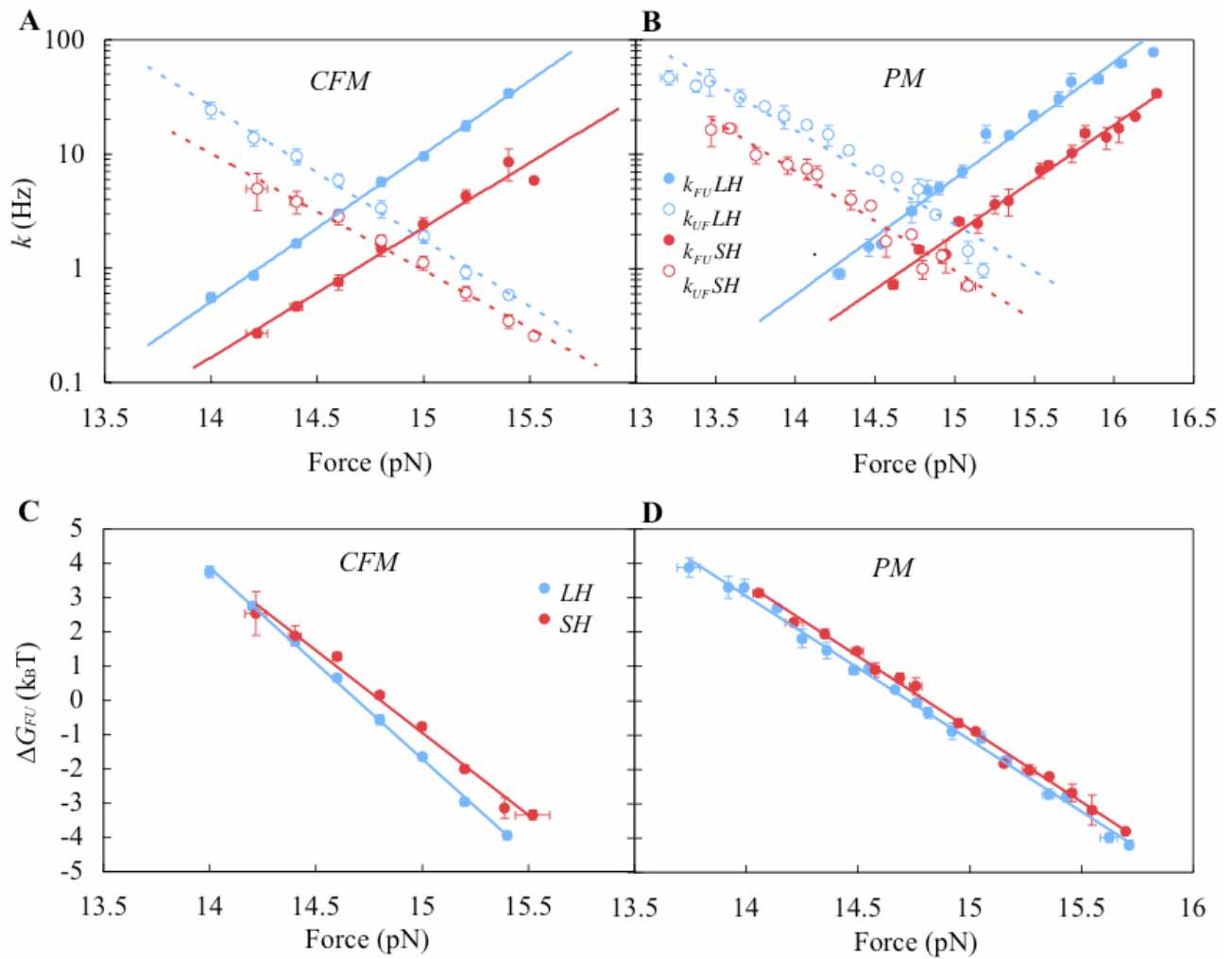

FIGURE 4 2S hairpin results with short handles (*red*) and long handles (*blue*). (*A*) Plots of $k$ as a function of force for the CFM experiments and (*B*) for the PM experiments (*solid circles* for $k_{FU}$ and *open circles* for $k_{UF}$), and the linear fit for the log of the rates (*solid lines* for $k_{FU}$ and *dotted lines* for $k_{UF}$). Plots of the $\Delta G_{FU}$ versus force for CFM (*C*) and PM (*D*) experiments and their linear fit (*solid lines*). We show the mean values for each point and the standard error. Statistics of molecules indicated in the caption of Table 1.

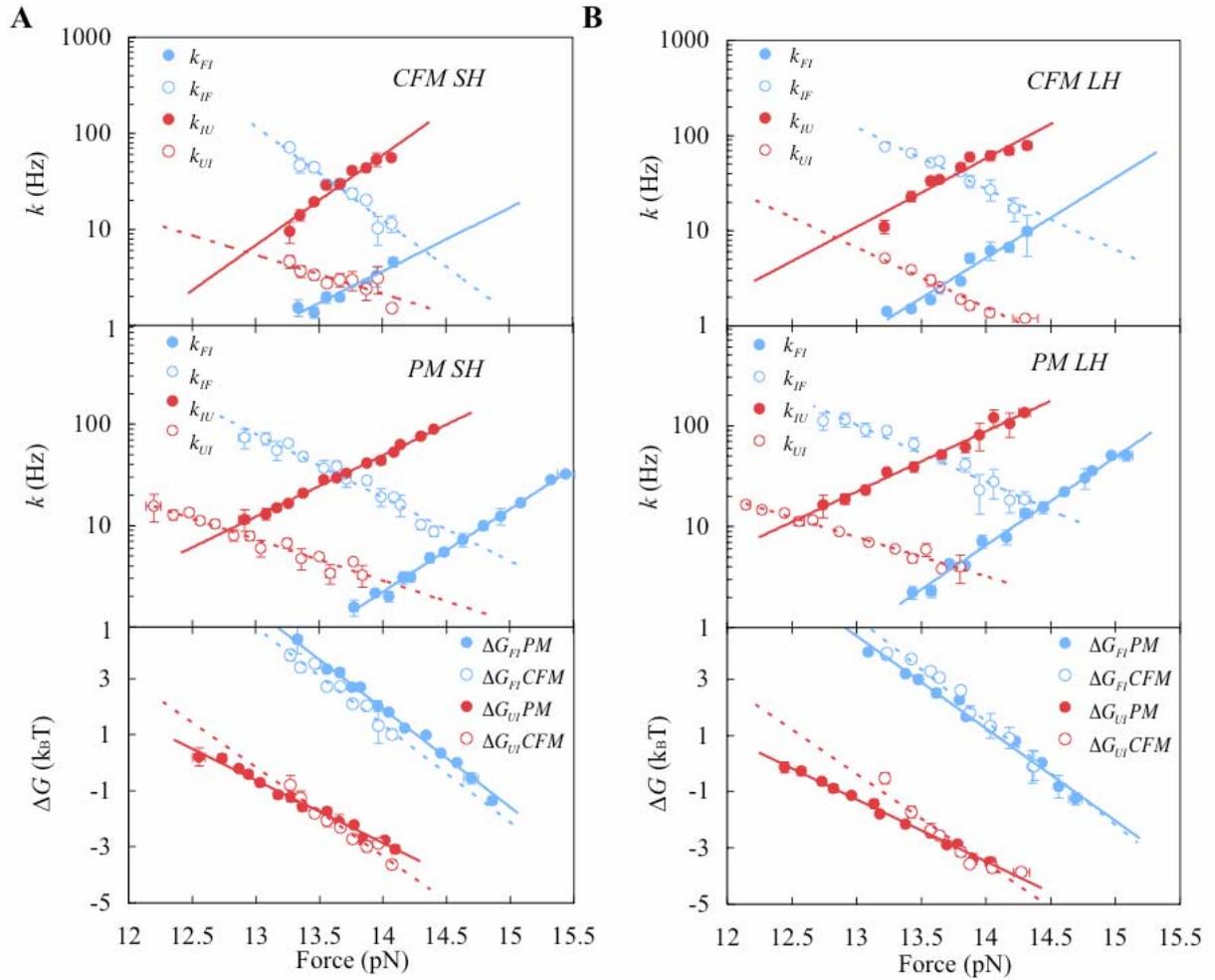

FIGURE 5 3S hairpin results. (*A*) Results for short handles and (*B*) long handles. Plots of $k$ as a function of force for the CFM (*upper panels*) and PM experiments (*middle panels*). Linear fits to the log of the rates are shown (*solid lines* for $k_{FI}$ and $k_{IU}$ and *dotted lines* for $k_{IF}$ and $k_{UI}$). *Bottom panels* show $\Delta G_{FI}$ versus force for PM (*blue solid circles*) and CFM (*blue open circles*) and $\Delta G_{UI}$ versus force for PM (*red solid circles*) and CFM (*red open circles*). The linear fits are shown as *solid lines* for PM, and *dotted lines* for CFM. We show the mean values for each point and the standard error. Statistics of molecules indicated in the caption of Table 1.

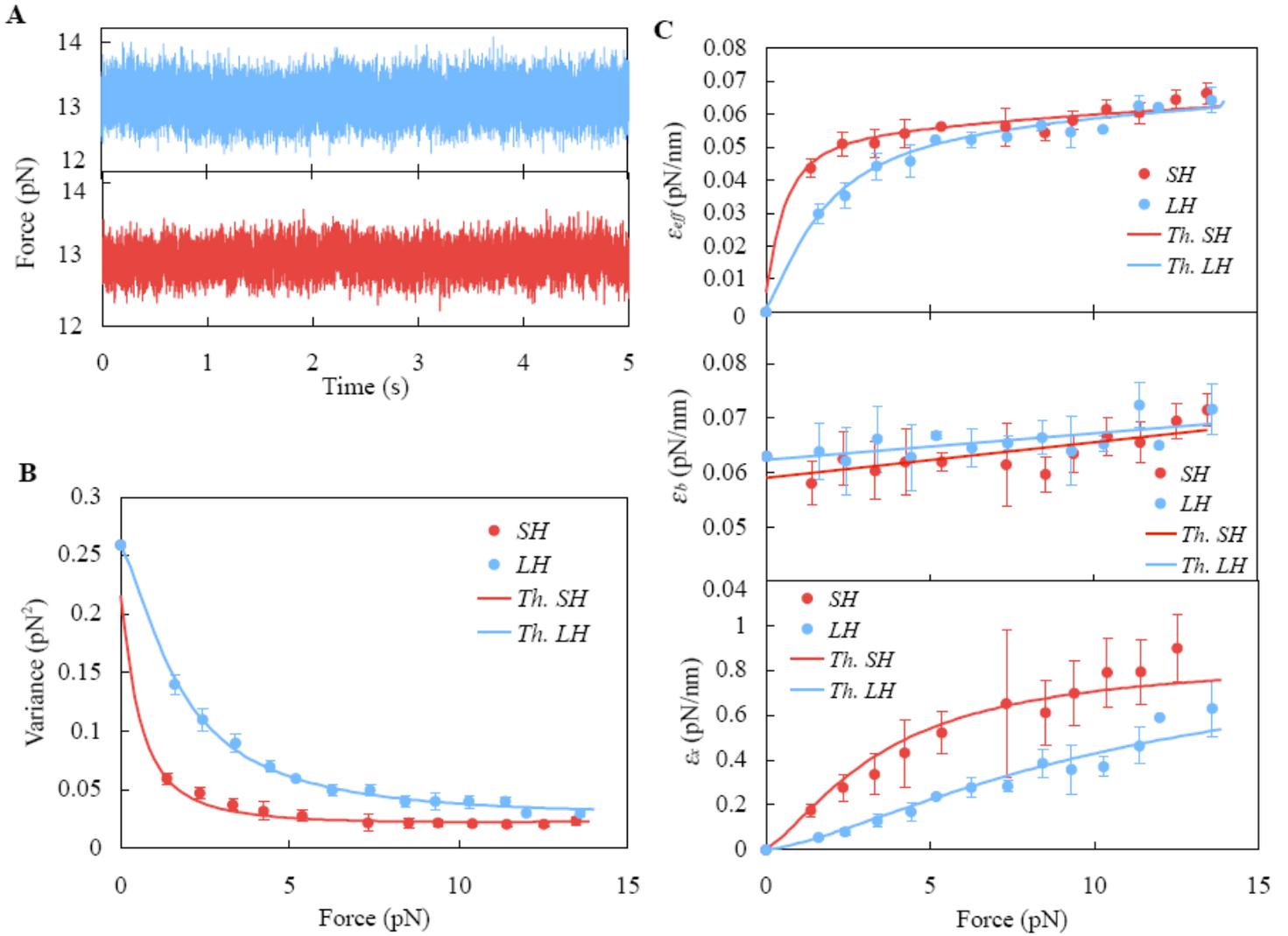

FIGURE 6 Analysis of force variance and stiffness (from 1 to 15 pN) for 2S hairpin. (*A*) Typical force traces (at ≅13 pN) with long (*blue*) and short handles (*red*). (*B*) Measured force variance for short (*red circles*) and long handles (*blue circles*). (*C*) The measured effective stiffness ($\varepsilon_{eff}$, *top panel*), the stiffness of the trap ($\varepsilon_b$, middle *panel*) and the stiffness of the molecular system ($\varepsilon_x$, *bottom panel*) for short (*red circles*) and long (*blue circles*) handles. Fits to the elastic model are shown for long (*blue line*) and short (*red line*) handles. Results are the average over 3 and 4 different molecules for the short and long handles cases respectively.

**SUPPORTING MATERIAL**

**S1. Free energy landscape and design of DNA hairpins**

**S2. Synthesis of DNA hairpins with short and long handles and molecular setup**

**S3. Experimental setup, hopping and pulling experiments**

**S4. Data analysis**

**S5. Folding free energy at zero force**

**S6. Rigidity of the optical trap**

**S7. Apparent rates at high and low trap stiffness**

**S8. Effect of the stretching modulus on the effective rigidity of an elastic polymer**

**S9. Full table of results**

## S1. Free energy landscape and design of DNA hairpins

The mechanical folding and unfolding of nucleic acid hairpins is commonly described in terms of a reaction coordinate and the corresponding free energy landscape (30-33) (see Fig. S1). When subject to force, the end-to-end distance of the molecule along the force axis is an adequate reaction coordinate for the folding-unfolding reaction pathway. For a given applied force $f$ it is common to consider only a single kinetic pathway for the unfolding and folding reactions, which is characterized by a single transition state (TS). The TS is the highest free energy state along the reaction coordinate and determines the kinetics of the folding-unfolding reaction. This model involves four parameters: the free energy difference between states S and S', $\Delta G_{SS'} = G_{S'} - G_S$; the height of the kinetic barrier $B$, defined as the free energy difference at force $f$ between the TS and the S state; and the distances $x^{\ddagger}_{SS'}$ and $x^{\ddagger}_{S'S}$ along the reaction coordinate that separates the TS from the S and S' states respectively. The total distance between S and S' is defined as $x_{SS'}$ ($x_{SS'} = x^{\ddagger}_{SS'} + x^{\ddagger}_{S'S}$). The distances and free energy differences between the different states (S, S' and TS) determine the force kinetics of unfolding/folding. Under an applied force the free energy landscape is tilted along the reaction coordinate changing the free energy difference $\Delta G_{SS'}$ and the barrier $B$. In a first approximation $\Delta G_{SS'}$ and $B$ change linearly with the force whereas $x^{\ddagger}_{SS'}$ and $x^{\ddagger}_{S'S}$, are taken as constant. This simplified representation can be generalized to include intermediates on-pathway.

In order to design the hairpins we have calculated the force-dependent molecular free energy landscapes by including the entropic contribution due to the stretching of the released ssDNA into the free energy landscape as predicted by the nearest-neighbor model (14). For the elastic response of the ssDNA we use the inextensible worm-like chain model (WLC) defined by,

$$F_{WLC}(x) = (k_B T/p)\left[\left(1/4(1-x/L_0)^2\right) - (1/4) + (x/L_0)\right], \quad (S1)$$

where $k_B$ is the Boltzmann constant and $T$ is the temperature, $p$ stands for the persistence length and $L_0$ is the contour length, $L_0 = n \times a$ where $a$=0.59 nm and $n$ is the total number of bases of the unfolded hairpin. For the persistence length we used values in the range 1.0-1.5nm that fit the experimentally measured force/distance jump at coexistence (see Section S5). Other models for the elastic behavior such as the freely jointed chain give very similar results.

To calculate the free energy landscapes shown in Fig. 2 *B* and Fig. 3 *B* we use the following formula,

$$G(n,f) = G_0(n) + G_{ssDNA}(0 \to x_n; f) - fx_n = G_0(n) - \int_0^f x_{FJC}(f')df', \quad (S2)$$

where we used $G_{ssDNA}(0 \to x_n; f) = \int_0^{x_n} F_{WLC}(x)dx$ with $f = F_{WLC}(x_n)$ as given in Eq. S1. Here $x_n$ stands for the molecular extension of *2n* bases of ssDNA (corresponding to the release of *n* bp of dsDNA) stretched at force *f*. The values for $G_0(n)$ can be obtained from Mfold (14,15) by adding the nearest-neighbor base pair free energies along the hairpin sequence. For *n=N* (where *N* stands for the total number of base pairs in the hairpin) the total number of bases is equal to *2n* plus the number of bases in the loop. Moreover, $G(n=N,f)$ must be corrected by adding the term $G_d(f)$ to account for the finite diameter $d_0$ of the hairpin which, in the presence of a force *f*, tends to be oriented along the force axis. The free energy cost to orient a dipole of length $d_0$ along an applied force *f* is given by,

$$G_d(f) = k_B T \ln[\sinh(\beta f d_0)/(\beta f d_0)], \quad (S3)$$

and the corresponding extension of the dipole is equal to

$$d(f) = d_0 \coth(\beta f d_0) - (1/\beta f), \quad (S4)$$

Note that in the limit *βfd₀>>1* the value of *d(f)* approaches $d_0$ whereas $G_d(f)$ is approximately equal to *fd₀*, the value typically employed in the literature to estimate the diameter contribution (38). For our calculations we took $d_0$=2 nm. This was the procedure we used for the 2S hairpin.

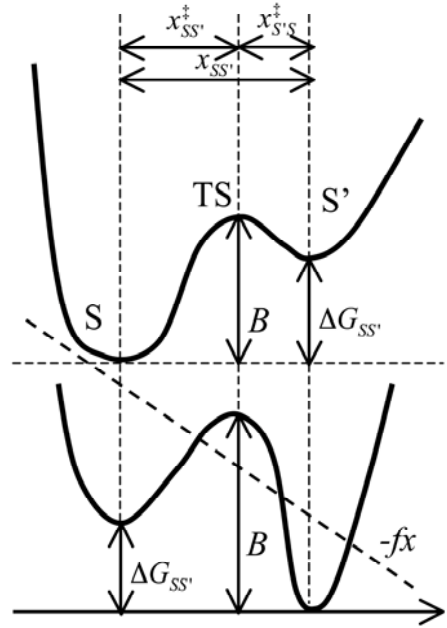

FIGURE S1 Schematic picture of the two-state model. The free energy landscape of the molecule along the reaction coordinate axis *x* at a given force has two minima corresponding to the two states *S* and *S'*. When a mechanical force is applied to the ends of the molecule the free energy landscape is tilted along *x*, decreasing de free energy of the *S'* state and the *TS* relative to the *S* state.

**S2. Synthesis of DNA hairpins with short and long handles and molecular setup**

The DNA hairpins (Section S1) with short handles are synthesized using the hybridization of three different oligonucleotides (Fig. 1 *B,* main text). One oligonucleotide contains the sequence of the ssDNA left handle plus a part of the sequence of the desired DNA hairpin; the second has the rest of the sequence of the DNA hairpin and the ssDNA right handle. The right and the left ssDNA handles have the same sequence to hybridize them with the third oligonucleotide. The first oligonucleotide has a biotin at its 5' end and the second oligonucleotide has been modified at its 3' end with a digoxigenin tail (DIG Oligonucleotide Tailing Kit, 2$^{nd}$ generation, Roche Applied Science, Barcelona, Spain). Once the first and the second oligonucleotides are hybridized to form the hairpin, the third oligonucleotide is hybridized to the handles to form the dsDNA handles of 29 bp each. All the oligonucleotide sequences used in this construction are shown in Fig. S2.

The 2S and 3S hairpins with long handles consist of a single DNA hairpin attached at its 5' and 3' end to long dsDNA handles used for pulling (Fig. S2 *A*). The left handle was synthesized through a PCR reaction using the pBR322 plasmid as a sample and the primers left-Biotin and left-Tsp45I (Fig. S2 *B*). The primer left-Biotin has a biotin at its 5' end. The product of the PCR was digested with the *Tsp*45I restriction enzyme (New England Biolabs, UK), giving a 528 bp dsDNA fragment with a biotin at one end and a nonpalindromic *Tsp*45I overhang at the other end. The right handle was obtained after two consecutive digestions of λ DNA. First, λ DNA is digested with *Sph*I enzyme (New England Biolabs, UK) and the 2216 bp fragment was gel purified. This fragment was digested again with *Tsp*RI (New England Biolabs, UK) and the 874 bp DNA gel purified. This dsDNA has at one end the cosL overhang of λ and at the other end a nonpalindromic *Tsp*RI sticky end. The cosL overhang was hybridized with the soc-Le oligonucleotide (Fig. S2 *B*) that was previously modified at 3' end with a digoxigenin tail (DIG Oligonucleotide Tailing Kit, 2$^{nd}$ generation, Roche Applied Science, Spain). The 2S and 3S DNA hairpins constructs are based on an oligonucleotide that has the desired sequence flanked by the two sticky ends (*Tsp*45I and *Tsp*RI) (Fig. S2 *B*). Finally, the hairpin was annealed and ligated to the two dsDNA handles to obtain the molecular construction.

Streptavidin-coated polystyrene microspheres (1.87 μm; Spherotech, Libertyville, IL) and protein G microspheres (3.0-3.4 μm; G. Kisker Gbr, Products for Biotechnologie, Steinfurt, Germany) coated with anti-digoxigenin polyclonal antibodies (Roche Applied Science, Spain) were used for specific attachments to the DNA molecular constructions described above. Attachment to the anti-digoxigenin microspheres was achieved first by incubating the beads with the tether DNA. The second attachment was achieved in the fluidics chamber and was accomplished by bringing a trapped anti-digoxigenin and streptavidin microspheres close to each other.

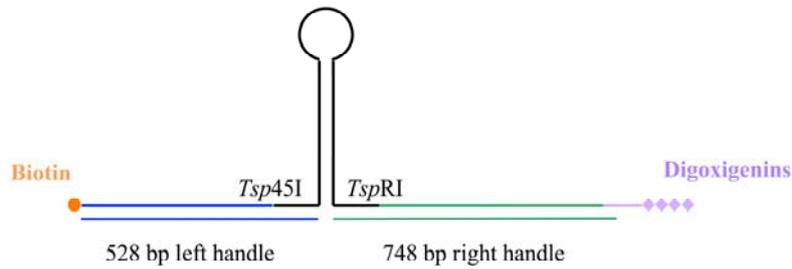

**A**

**B**

| Primers PCR left handle |
|---|
| TTC TTG AAG ACG AAA GGG CCT |
| CCATTGCTGCAGGCATCGTG |
| **soc-Le oligonucleotide** |
| AGGTCGCCGCCCAAAAAAAAAAAA |
| **2S hairpin** |
| **GTCAC**GCGAGCCATAATCTCATCTGGAAACAGATGAGATTATGGCTCGC**GGCAGTGTT** |
| **3S hairpin** |
| **GTCAC**GCGTCGCAGCGCCAAAAGGCAGGCGGAAAGAGCGCCTGCCTTTTCGCTGCGACGC**GGCAGTGTT** |

**C**

| | 2S hairpin |
|---|---|
| 1st oligo | **AGTTAGTGGTGGAAACACAGTGCCAG**CGCGCGAGCCATAAT |
| 2nd oligo | CTCATCTGGAAACAGATGAGATTATGGCTCGC**AGTTAGTGGTGGAAACACAGTGCCAGCGC** |
| | **3S hairpin** |
| 1st oligo | **AGTTAGTGGTGGAAACACAGTGCCAG**CGCGCGTCGCAGCGCCAAAAGGCAGGCGGAAAGAGCGCCTGCCTTTTCGCTG |
| 2nd oligo | CGACGC**AGTTAGTGGTGGAAACACAGTGCCAGCGC** |
| | **3rd oligo** |
| | GCGCTGGCACTGTGTTTCCACCACTAACT |

FIGURE S2 (*A*) Molecular construct with dsDNA long handles, left handle (528bp) made by PCR reaction and right handle (748 bp) obtained from λ DNA. (*B*) In the table are listed the oligonucleotides used to make the long handles construction. The 2S and 3S hairpins sequences have in bold the two sticky ends (*Tsp*45I and *Tsp*RI). (*C*) Sequences used to synthesize the two hairpins with short handles. In bold is shown the part of the sequence that corresponds to the handle.

**S3. Experimental setup, hopping and pulling experiments**

The experiments have been carried out using a high stability newly designed miniaturized dual-beam optical tweezers apparatus (32,37). It consists of two counter-propagating laser beams of 845 nm wavelength that form a single optical trap where particles can be trapped by gradient forces. The DNA hairpin is tethered between two beads (Fig. 1 *A*). One bead is immobilized in the tip of a micropipette that is solidary with the fluidics chamber; the optical trap captures the other bead. The light deflected by the bead is collected by two photodetectors located at opposite sides of the chamber that produce a direct measure of the total change in light momentum which is equal to the net force acting on the bead. Piezo actuators coupled to metallic wigglers that bend the optical fibers can move the optical trap.

The folding-unfolding experiments described in this report were performed at ambient temperature (25ºC) in a buffer containing 10 mM Tris-HCl pH7.5, 1 mM EDTA, 1 M NaCl, 0.01% Sodium Azide. Two types of hopping experiments were done for the DNA hairpin constructs:

1. CFM (33,39,40): the force applied to the DNA constructs was maintained to a preset value (usually between 12 and 15 pN) by moving the piezo actuators through a feedback control (33) that operates at 1kHz. We can observe the molecule hopping between different extensions depending on the state of the hairpin (Fig 2F and Fig 3F main text).

2. PM (39,40): in this case the position of the trap is kept constant (without feedback) and allowing the captured bead to passively move in the optical trap. Thus the trapped bead changes position inside the trap in response to the end-to-end distance change of the molecular construct. Consequently, the force hops among different levels corresponding to the different states of the hairpin (Fig 2D and Fig 3D main text). By moving the trap to a new position, the value of the tension on the hairpin in the different (folded, unfolded and intermediate) states changes, modifying the equilibrium (Boltzmann-Gibbs) weights of these states. This kind of experiment allowed us to measure the kinetic rates over different forces.

In pulling experiments the optical trap is moved at a constant speed and the molecule pulled (3,16) until a value of the force is reached such that the molecule unfolds. If the pulling process is reversed then the molecule refolds again. In these experiments the force exerted upon the system is recorded as the function of the trap relative distance giving the so-called force-distance curve (FDC). The folding and refolding of the molecule can be identified as force-distance jumps observed in the FDC. In all cases data were collected at 1 kHz.

**S4. Data analysis**

Detailed balance condition Eq. 2 can also be applied to the PM case, where the control parameter is the position of the trap relative to the pipette, $X_T$, rather than the force. In that latter case we can expand the free energy difference between states S and S' in the vicinity of the coexistence distance $X_T$ as: $\Delta G_{SS'}(X_T) = -(X_T - X_T^c)\Delta f$, where $\Delta f$ is the (positive) force jump between the folded and unfolded branches. This expression tells us that at coexistence ($X_T = X_T^c$) the two states have the same free energy and, along each branch, we have $f = \partial G/\partial X_T$. By using the relation $\Delta f = \varepsilon_{eff} x_m$ with $\varepsilon_{eff}$ and $x_m$ the effective stiffness and the molecular extension of ssDNA released at the transition we get, $\Delta G_{SS'}(\overline{f}) = -(\overline{f} - \overline{f}_c)\frac{\Delta f}{\varepsilon_{eff}} = -(\overline{f} - \overline{f}_c)x_m = \Delta G_{SS'} - \overline{f}x_m$ where $\overline{f}$ is average force between the folded and the unfolded branches at a given $X_T$ and $\Delta G_{SS'} = \overline{f}^c x_m$ is the free energy of formation at zero force plus the stretching contribution of the extended ssDNA. Therefore Eq. 2 can be also used in the PM case replacing the force $f$ by the average force between the folded and the unfolded branches, $\overline{f}$.

For the PM hopping experiments of the two-states 2S hairpin, folding and unfolding transition rates were calculated from the time-dependent force traces (39) (Fig. 2*D*). Each trace normally contained 50-150 cycles of unfolding/refolding events, which showed no significant force drift. Distributions of the force were fitted to Gaussian functions for the folding and unfolding processes (examples are shown in Fig. 2*E*),

$$P(f) = \left(w/(2\pi\sigma_F^2)\right)^{1/2} e^{-\left((f-f^F)^2/2\sigma_F^2\right)} + \left(1-w/(2\pi\sigma_U^2)\right)^{1/2} e^{-\left((f-f^U)^2/2\sigma_U^2\right)},$$

(S5)

where $P(f)$ is the normalized number of counts for each binned force $f$; $w$ and $1-w$ are the statistical weights of the unfolded and folded states and $\sigma_n^2$ (n = U or F) are the widths of the Gaussian peaks, respectively; $f^U$ and $f^F$ are the average forces at the unfolded and folded states, respectively; States (folded or unfolded) of the hairpin along the force trace were assigned according to whether the instantaneous force was closer to $f^U$ or $f^F$. Transition rates were computed as the inverse of the mean lifetime for each state. From Eq. 1a and Eq. 1b and doing linear fits of the logarithm of the rates versus force, we can extract the free energy difference between states F and U by using the expression, $\Delta G_{FU} = k_B T [\log(k_{U \to F}) - \log(k_{F \to U}) + fx_{FU}]$, but using the Eq. 2 we can also obtain these values.

For the 3S hairpin PM hopping (see Fig. 3*D* for an experimental trace) we applied the same data analysis as for the 2S hairpin but including the intermediate state, I. We assume that to go from F to U and vice versa the hairpin always goes through I; therefore, four different transition rates were obtained: transition rates from F to I, from I to F, from I to U and finally from U to I. Distributions of the force were fitted to three Gaussian functions for the F, I and U states (Fig. 3*E*),

$$P(f) = \left(w_U / (2\pi\sigma_U^2)^{1/2}\right) e^{-\left((f-f^U)^2 / 2\sigma_U^2\right)} + \left(w_F / (2\pi\sigma_F^2)^{1/2}\right) e^{-\left((f-f^F)^2 / 2\sigma_F^2\right)}$$
$$+ \left((1-w_F-w_U) / (2\pi\sigma_I^2)^{1/2}\right) e^{-\left((f-f^I)^2 / 2\sigma_I^2\right)},$$

(S6)

where *P(f)* is the normalized number of counts for each binned force *f*; $w_U$, $w_F$ and $w_I = 1-w_F-w_U$ are the statistical weights of the U, F and I states and $\sigma_n^2$ (n = U, F or I) are the widths of the Gaussian peaks, respectively; $f^U$, $f^F$ and $f^I$ are the average forces at the U, F and I states, respectively; States (F, U and I) of the DNA along the force trace were assigned according to whether the instantaneous force was closer to $f^U$, $f^F$ and $f^I$. Transition rates were computed as the inverse of the mean lifetime for each state. The free energy difference $\Delta G_{SS'}$ between any pair of states S, S´ (F, U and I) is obtained from the rates Eq. 1a and Eq. 1b and using the Eq. 2 as we did for the hairpin 2S.

For the CFM, transition rates and free energy differences were calculated from the time-dependent extension traces (Fig. 2 *F* and Fig. 3 *F*). Hopping traces usually contain 50-150 folding/refolding cycles. As the measured extension traces may drift over the time period, we applied a different strategy to analyze these data. A transition between the F and U states in the 2S hairpin was considered to occur when the extension changed by at least 60 % of the average total extension difference between both states, and by at least 50 % of the extension difference between the F and I states and between the I and U states for the 3S hairpin.

**S5. Folding free energy at zero force**

From hopping experiments we have determined the different values of $\Delta G_{SS'}$ for the different hairpins (Table 1). To extract the free energy difference between different states at zero force ($\Delta G^0_{SS'}$) we must subtract to the experimentally determined $\Delta G_{SS'}$ from Eq. 1a, Eq. 1b and Eq. 2 the contribution of mechanical stretching of the ssDNA at the coexistence force (6) as well as the orientation of the hairpin. The most straightforward way of doing this is by measuring the value of the coexistence force and the released molecular extension and to use the WLC model with parameters previously given (see Section S1),

$$\Delta G^0_{SS'} = \int_0^{f^C_{SS'}} x_{WLC}(f')df' - k_B T \ln\left[\sinh(\beta f^C_{SS'} d_0)/(\beta f^C_{SS'} d_0)\right], \tag{S7}$$

where $f^C_{SS'}$ stands for the coexistence force between states S and S' and $d_0$ is the diameter of the hairpin (taken equal to 2 nm). The second term in the rhs of Eq. S7 corresponds to the free energy correction due to the orientation of the hairpin along the force axis (Section S1 and Eq. S3). Let us stress that previous expression Eq. S7 does not require to include the free energy correction expected from the contraction/expansion of handles or the repositioning of the bead when the molecule hops.

How do we estimate the value of $\Delta G^0_{SS'}$? We proceed as follows. In the CFM handles and bead contraction are not expected because the force is kept constant. Therefore we used the following expression,

$$x_{WLC}(f^0_{SS'}) = \Delta x_{SS'} + d(f^0_{SS'}), \tag{S8}$$

where $\Delta x_{SS'}$ is the experimentally measured average molecular extension jump between the states S and S' along the hopping trace at coexistence and $d(f^0_{SS'})$ is the average extension contributed by the orientation of the hairpin as given in Eq. S4. We then determined the persistence length value for the ssDNA such that Eq. S8 holds. Using these values in Eq. S7 we could then determine the value of $\Delta G^0_{SS'}$.

In PM experiments the change in force when the molecule hops induces changes in the molecular extension of the handle and the bead position. In this case it is easy to prove (see Appendix C in (38)) that

$$x_{WLC}(f^C_{SS'}) = \left(\Delta f_{SS'}/\varepsilon_{\it eff}(f^C_{SS'})\right) + d(f^C_{SS'}), \tag{S9}$$

where $\Delta f_{SS'}$ is the average force jump between the states S and S' along the hopping trace at coexistence and $\varepsilon_{eff}(f_{SS'}^C)$ is the effective rigidity of the molecular setup (bead and handles) in the high force state (might be S or S') at coexistence. In order to extract the folding free energy at zero force we determined the value of the persistence length for the ssDNA such that released extension agrees with the experimental estimate obtained from Eq. S9. The same procedure was used in the CFM, obtaining identical elastic ssDNA parameters. Using these values in Eq. S7 we could then determine the value of $\Delta G_{SS'}^0$.

In all conditions we investigated (DNA sequence, PM versus CFM hopping, long handles versus short handles) the values obtained for $\Delta G_{SS'}^0$ agree reasonably well with the values estimated from the unified oligonucleotide parameters used by Mfold servers to predict folding free energies (14). However, as it is well known, this is strongly dependent on the model used to describe the ideal elastic properties of the ssDNA.

**TABLE S5 Folding free energy at zero force of 2S and 3S hairpins.**

|    | $x_{FU}$ | $f_{FU}^C$ | $G_d$ | $G_{ssDNA}$ | $\Delta G_{FU}^0$ | $\Delta G_{MFold}^0$ |
|---|---|---|---|---|---|---|
| **2S** | 18.3 ±0.9 | 14.8 ±0.7 | 1.7 | 16.6 ±0.2 | 50.9 ±0.7 | 50.7 |
|    | $x_{FI}$ | $f_{FI}^C$ | $G_d$ | $G_{FI}^{ssDNA}$ | $\Delta G_{FI}^0$ | $\Delta G_{MFold}^{FI}$ |
| **3S** | 12.0 ±0.6 | 14.5 ± 0.7 | - | 9.7 ± 0.2 | 32.6 ±3.5 | 33.9 |
|    | $x_{IU}$ | $f_{IU}^C$ | $G_d$ | $G_{IU}^{ssDNA}$ | $\Delta G_{IU}^0$ | $\Delta G_{MFold}^{IU}$ |
| **3S** | 10.0 ±0.7 | 12.8 ±0.6 | 1.5 | 9.9 ±0.1 | 22.7 ±3.3 | 27.2 |

Results of the 2S hairpin in the first row and the results for 3S hairpin in the two last rows. To measure the folding free energies at zero force we used the coexistence forces and extensions given in Table 1. The forces are given in pN, the extensions in nm and the energies in $k_BT$. Stretching contributions were estimated using the elastic parameters reported in Section S5. For the 2S hairpin we took a total number of bases $N$ equal to 44 and hairpin diameter $d=2$ nm. For the 3S hairpin we took $N=26$ and $d=0$ between states F and I and $N=29$ and $d=2$ nm between states I and U. Uncertainties in free energies were estimated by propagating the experimental errors obtained for the values of the molecular extension and the coexistence force. Statistics of molecules indicated in the caption of Table 1.

**S6. Rigidity of the optical trap**

The measurement of the power spectrum is a calibration method that uses the thermal fluctuations of a bead in the optical trap to determine the stiffness of the trap. The experiments have been carried out with calibration beads of 3 μm of diameter in the same buffer where we did the experiments with hairpins. The force fluctuations have been measured at an acquisition rate of 50 kHz using a data acquisition board (National Instrument PXI-1033), which allows us to achieve a wide range of frequencies. The power spectral density has been calculated from 500000 data points. Fig. S6 shows the power spectrum obtained by the Fourier transform of the experimental data. The power spectrum has been fitted to a theoretical Lorentzian function,

$$\langle \Delta f^2(\nu) \rangle = \left(2\xi k_B T \omega_c^2\right) / (\omega_c^2 + (2\pi\nu)^2) = S^2 \left(a/(b + (2\pi\nu)^2)\right), \tag{S10}$$

where $\nu$ is the frequency in Hz, $\omega_c$ is the corner frequency in rad/s ($\omega_c = 2\pi\nu_c$), $\xi$ corresponds to the drag coefficient, $S$ is the conversion factor from Volts to pN, and $a$ and $b$ are the fitting parameters of the Lorentzian function.

By fitting Eq. S10 to the power spectrum and from the knowledge of the value of the drag coefficient we can determine the conversion factor $S$ and $\omega_c$. The value of $\xi$ has been obtained from a measurement of the viscosity ($\eta$) of the buffer in a viscosimeter and using the relation $\xi = 6\pi\eta r$ with $r$=1.5 μm (the radius of the calibration bead). We find $\xi = 2.78 \times 10^{-5}$ pN s/nm, $S \cong 23$ pN/V S and $\omega_c$=2293 Hz. The stiffness of the optical trap $\varepsilon_b$ is obtained with the equation:

$$\varepsilon_b = \omega_c \xi, \tag{S11}$$

where we obtain a stiffness value of 0.064 pN/nm.

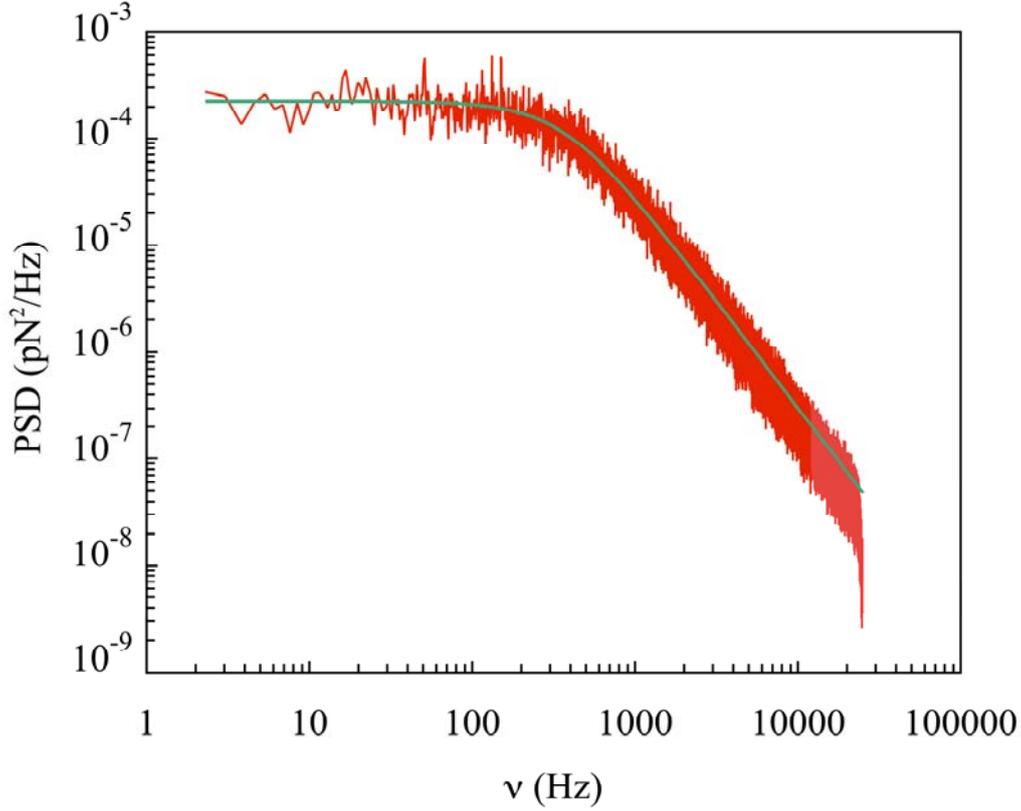

FIGURE S6 Calibration by thermal noise. Power spectrum of the Fourier transform of the fluctuation force data of the bead in the optical trap (in *red*) and the Lorentzian fit (in *green*)

**S7. Apparent rates versus passive mode rates at high and low trap stiffness**

Under PM conditions the folding/unfolding rates can be plotted or represented in two different ways: as a function of the average force in the folded and unfolded states (this is the standard representation adopted throughout this paper) or as a function of the trap position (the true control parameter in PM experiments). The apparent coexistence rate $k_{app}^C$ is the value of the folding/unfolding transition rate in the latter representation, where both states (F and U) are equally populated. Note that $k_{app}^C$ differs from the PM coexistence rates, $k_{FU}^C$, also measured in the same PM experiments. Because the PM coexistence rates have been found to be nearly equal to the coexistence rates in the CFM, they seem a more robust indicator about force kinetics than the apparent coexistence rates (39,40). This is the reason why we adopted PM rates throughout this paper. According to (41) (see the Supplementary Material shown in (40) as well), the apparent coexistence rate $k_{app}^C$ should decrease when the trap stiffness decreases and should be larger than the CFM or PM coexistence rates $k_{FU}^C$ (obtained from the crossing points of the linear fits of the $k_{FU}$ and $k_{UF}$) (39,40). Similarly one might expect that the apparent coexistence rate $k_{app}^C$ should decrease for longer handles as compared to short handles. Strikingly enough, this is the contrary of what we find for the PM coexistence rates: they tend to increase for longer handles, i.e. when the effective rigidity of the molecular setup decreases. Is there a discrepancy between our results and those reported in (41)? We have challenged this apparent contradiction by doing PM hopping experiments at high (0.064 pN/nm) and at low (0.020 pN/nm) trap stiffness with the 2S hairpin with short and long handles (see Fig. S7 and Table S7). As expected we have found that $k_{app}^C$ is always higher than the PM coexistence rate. Our results also confirm the striking dependences reported in this and previous works: although PM coexistence rates increase either by decreasing the power of the trap or increasing the length of the handles, the values of $k_{app}^C$ follow the reverse tendency and increase as the effective rigidity of the setup is decreased.

Note that the difference in kinetics between apparent and coexistence rates for short and long handles can be explained in terms of the force jump measured in the passive mode which is proportional to the effective stiffness of the setup formed by the trap serially connected to the handles. Because the stiffness of the handles (short and long) that are pulled at approximately 14pN is much higher than the stiffness of the trap, the effective stiffness is approximately equal to the stiffness of the trap. In Figure S7-2 we show an illustration of such effect. Consequently, the ratio $k_{app}^C / k_{FU}^C$ is ~ 3.2 for both long and short handles at 0.06 pN/nm, and ~1.5 for both handle lengths at 0.02 pN/nm, i.e. to a very good approximation it is only dependent on the trap stiffness.

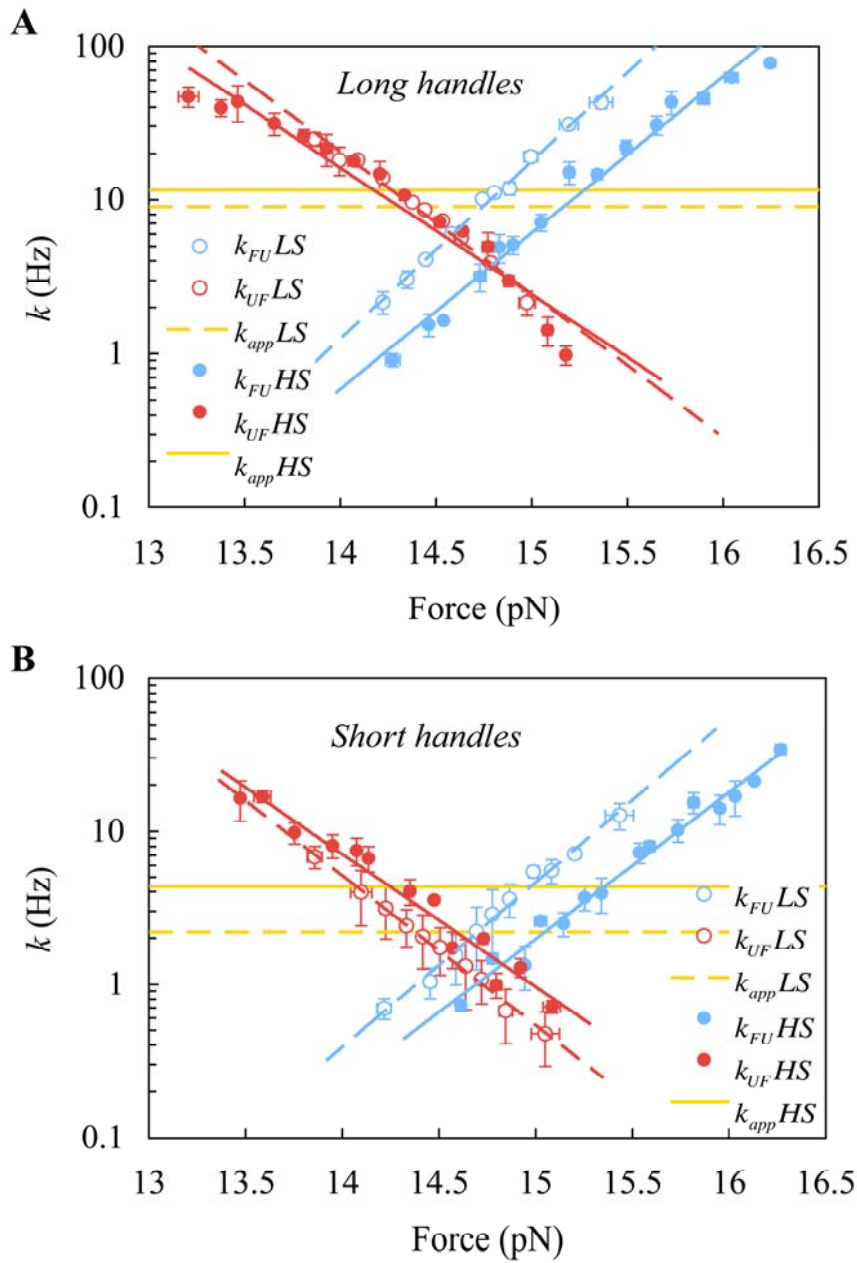

FIGURE S7 Apparent coexistence rates for 2S hairpin with long handles. (*A*) and short handles (*B*). Plots of the *k* as a function of force for PM experiments done at low trap stiffness (*open circles*) and at high trap stiffness (*solid circles*) and their linear fit (*dotted lines* for low trap stiffness and *solid lines* for high trap stiffness). The $k_{FU}$ is shown in *blue* and the $k_{UF}$ is shown in *red*. The apparent coexistence rates are shown for low (*yellow dotted line*) and high trap stiffness (*yellow solid line*). The values are obtained averaging 2 molecules for low trap stiffness and 5 and 7 molecules for high stiffness with short and long handles respectively.

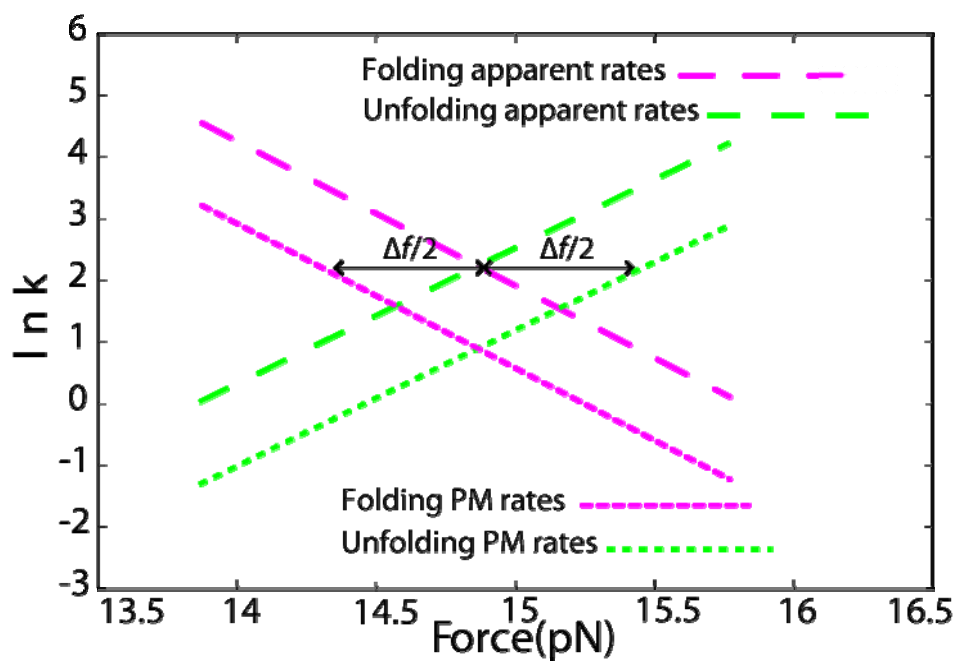

FIGURE S7-2: Schematics of the logarithm of the kinetic rates versus force representation for the apparent rates and PM rates measured in the PM. The switch between the lines correspond to the jump in force $\Delta f$ between folded and unfolded branches.

**TABLE S7 Coexistence rates of 2S hairpin with long and short handles**

|  | Long handles | | Short handles | |
| --- | --- | --- | --- | --- |
|  | 0.064 pN/nm | 0.020 pN/nm | 0.064 pN/nm | 0.020 pN/nm |
| $k_{app}^{C}$ | 11.7 ±0.6 | 9.07 ±0.06 | 4.4 ±0.4 | 2.2 ±0.7 |
| $k_{FU}^{C}$ | 3.8 ±0.4 | 5.9 ±0.2 | 1.3 ±0.2 | 1.4 ±0.4 |

The rates are given in Hz. The values are obtained averaging the results of 2 molecules for low trap stiffness and 5 and 7 molecules for high stiffness with short and long handles respectively. For each cell the top value is the average and the bottom value is the standard error.

**S8. Effect of the stretching modulus on the effective rigidity of an elastic polymer**

In this section we address the question of how the extensibility of a polymer modifies the rigidity associated to the entropic elasticity. Let us consider an inextensible polymer characterized by its entropic elasticity, $f_{L_0}(x) = \hat{f}(x/L_0)$ where $x$ is the molecular extension and $L_0$ is the contour length. The corresponding rigidity of the inextensible polymer is given by $\varepsilon_i = (1/L_0) \times (\hat{f})'(x/L_0)$, meaning that, at a given force $f$, the rigidity of the handle $\varepsilon_h$ is inversely proportional to the contour length. The extensibility of the polymer can be modeled by assuming that the contour length $L_0$ changes with force according to $L_0(1+f/Y)$ where $f$ is the force and $Y$ stands for the stretching modulus. It is straightforward to prove that the rigidity $\varepsilon_e$ of the extensible polymer can be written as,

$$1/\varepsilon_e = 1/\varepsilon_i + x/Y, \tag{S12}$$

showing that only when $x \gg Y/\varepsilon_i$ the contribution of the extensibility to the rigidity of the inextensible polymer is important. For dsDNA its elastic properties are well described by the worm-like chain model,

$$f = (k_B T/p) \times \left[ (1/4(1-x/L_0)^2) - (1/4) + (x/L_0) \right], \tag{S13}$$

where $p$ stands for the persistence length. Extensibility affects the rigidity of the polymer only when $x \cong L_0$ (i.e. when $\varepsilon_i$ is maximum). In such limit we can approximate Eq. S13 by $f \approx (k_B T/4p)(1-x/L_0)^{-2}$ which gives

$$\varepsilon_i \approx k_B T/2pL_0 \left[ (4pf/k_B T)^{3/2} + 2 \right] \approx (4f^{3/2}/L_0)(p/k_B T)^{1/2}, \tag{S14}$$

This expression can be introduced into Eq. S12 yielding,

$$1/\varepsilon_e \approx \left[ (k_B T/p)^{1/2} (1/4 f^{3/2}) + (1/Y) \right] L_0, \tag{S15}$$

For a linear dsDNA with $p \cong 50$nm and $Y \cong 1000$pN, the contribution of the extensibility property (i.e. the term $1/Y$ in Eq. S15) to the effective rigidity of the polymer starts to be important at forces above 10 pN. Equation S12 is a valid interpolation over a wide range of forces (above $\cong 1$ pN). Note that for $f \gg 20$pN, $\varepsilon_e$ is constant and given approximately by $Y/L_0$.

## S9. Full table of results

Errors in this table are only statistical and do not include systematic instrumental uncertainties due to force and distance calibration errors (full errors combining statistical and instrumental uncertainties are given in Table 1 of main text and Table S5).

*Table S9a. Kinetic and thermodynamic parameters for the 2S hairpin with short handles.*

| 2S SH PM BELL - EVANS | $f^C_{FU}$ | $k^C_{FU}$ | $x^{\ddagger}_{FU}$ | $x^{\ddagger}_{UF}$ | $x_{FU}$ | $\Delta G_{FU}$ |
|---|---|---|---|---|---|---|
| Mol1 | 14.71 | 1.47 | 9.79 | 8.79 | 18.58 | 66.48 |
| Mol2 | 14.95 | 0.71 | 9.45 | 8.64 | 18.09 | 65.78 |
| Mol4 | 15.19 | 0.96 | 9.63 | 7.84 | 17.47 | 64.56 |
| Mol5 | 14.89 | 1.63 | 8.53 | 8.01 | 16.54 | 59.91 |
| Mol6 | 14.45 | 1.73 | 9.15 | 8.15 | 17.30 | 60.81 |
| **Average** | **14.8** | **1.30** | **9.31** | **8.28** | **17.59** | **63.5** |
| Statistical error | 0.12 | 0.20 | 0.22 | 0.18 | 0.35 | 1.33 |

| 2S SH CFM BELL - EVANS | $f^C_{FU}$ | $k^C_{FU}$ | $x^{\ddagger}_{FU}$ | $x^{\ddagger}_{UF}$ | $x_{FU}$ | $\Delta G_{FU}$ |
|---|---|---|---|---|---|---|
| Mol1 | 14.85 | 1.76 | 9.88 | 11.44 | 21.31 | 76.99 |
| Mol4 | 15.19 | 0.97 | 9.79 | 9.33 | 19.12 | 70.64 |
| Mol5 | 14.86 | 1.57 | 11.90 | 10.08 | 21.99 | 79.51 |
| Mol6 | 14.46 | 1.66 | 9.80 | 9.47 | 19.27 | 67.78 |
| **Average** | **14.8** | **1.49** | **10.3** | **10.1** | **20.42** | **73.7** |
| Statistical error | 0.15 | 0.18 | 0.52 | 0.48 | 0.72 | 2.72 |

| 2S SH PM DETAILED BALANCE | $f^C_{FU}$ | $x_{FU}$ | $\Delta G_{FU}$ |
|---|---|---|---|
| Mol1 | 14.68 | 18.59 | 66.39 |
| Mol2 | 14.92 | 18.16 | 65.91 |
| Mol4 | 15.13 | 17.52 | 64.52 |
| Mol5 | 14.87 | 16.55 | 59.88 |
| Mol6 | 14.42 | 17.28 | 60.62 |
| **Average** | **14.8** | **17.59** | **63.5** |
| Statistical error | 0.12 | 0.35 | 1.35 |

| 2S SH CFM DETAILED BALANCE | $f^C_{FU}$ | $x_{FU}$ | $\Delta G_{FU}$ |
|---|---|---|---|
| Mol1 | 14.83 | 25.93 | 93.56 |
| Mol4 | 15.15 | 18.26 | 67.27 |
| Mol5 | 14.85 | 22.26 | 80.46 |
| Mol6 | 14.45 | 18.87 | 66.34 |
| **Average** | **14.8** | **21.3** | **76.9** |
| Statistical error | 0.14 | 1.8 | 6.42 |

*Table S9b. Kinetic and thermodynamic parameters for the 2S hairpin with long handles.*

| 2S LH PM BELL – EVANS | $f_{FU}^C$ | $k_{FU}^C$ | $x_{FU}^{\ddagger}$ | $x_{UF}^{\ddagger}$ | $x_{FU}$ | $\Delta G_{FU}$ |
|---|---|---|---|---|---|---|
| Mol1 | 15.43 | 5.32 | 9.94 | 7.44 | 17.38 | 64.85 |
| Mol2 | 14.85 | 2.91 | 9.95 | 8.32 | 18.27 | 67.87 |
| Mol3 | 14.63 | 4.09 | 9.7 | 9.54 | 19.24 | 68.49 |
| Mol4 | 14.58 | 2.66 | 9.98 | 9.19 | 19.17 | 70.88 |
| Mol5 | 14.68 | 3.57 | 10.27 | 8.58 | 18.85 | 67.33 |
| Mol6 | 14.74 | 3.36 | 9.85 | 8.74 | 18.59 | 66.66 |
| **Average** | **14.78** | **3.79** | **9.65** | **8.70** | **18.35** | **66.2** |
| Statistical error | 0.12 | 0.36 | 0.30 | 0.26 | 0.33 | 1.20 |

| 2S LH CFM BELL – EVANS | $f_{FU}^C$ | $k_{FU}^C$ | $x_{FU}^{\ddagger}$ | $x_{UF}^{\ddagger}$ | $x_{FU}$ | $\Delta G_{FU}$ |
|---|---|---|---|---|---|---|
| Mol2 | 14.86 | 3.6 | 11.96 | 10.41 | 22.37 | 80.48 |
| Mol3 | 14.54 | 5.27 | 13.04 | 11.03 | 24.07 | 85.88 |
| Mol4 | 14.67 | 3.06 | 12.13 | 10.03 | 22.16 | 79.53 |
| Mol5 | 14.67 | 4.2 | 11.87 | 12.09 | 23.96 | 85.71 |
| Mol6 | 14.73 | 3.83 | 11.83 | 11.08 | 22.91 | 82.67 |
| **Average** | **14.69** | **3.99** | **12.17** | **10.93** | **23.09** | **82.85** |
| Statistical error | 0.05 | 0.37 | 0.22 | 0.35 | 0.40 | 1.30 |

| 2S LH PM DETAILED BALANCE | $f_{FU}^C$ | $x_{FU}$ | $\Delta G_{FU}$ |
|---|---|---|---|
| Mol1 | 14.52 | 18.66 | 65.93 |
| Mol2 | 14.81 | 17.71 | 63.79 |
| Mol3 | 14.63 | 18.31 | 65.16 |
| Mol4 | 14.56 | 18.85 | 66.77 |
| Mol5 | 14.64 | 18.33 | 65.2 |
| Mol6 | 14.72 | 18.03 | 64.55 |
| Mol7 | 14.57 | 16.89 | 59.97 |
| **Average** | **14.63** | **18.11** | **64.49** |
| Statistical error | 0.04 | 0.25 | 0.83 |

| 2S LH CFM DETAILED BALANCE | $f_{FU}^C$ | $x_{FU}$ | $\Delta G_{FU}$ |
|---|---|---|---|
| Mol2 | 14.86 | 22.52 | 81.44 |
| Mol3 | 14.54 | 24.11 | 85.31 |
| Mol4 | 14.67 | 22.22 | 79.32 |
| Mol5 | 14.69 | 22.75 | 81.30 |
| Mol6 | 14.71 | 23.08 | 82.63 |
| **Average** | **14.69** | **22.94** | **82.00** |
| Statistical error | 0.051 | 0.32 | 0.98 |

Table S9c. *Kinetic and thermodynamic parameters for the 3S hairpin with short handles.*

| 3S SH PM Bell-Evans | $f_{FI}^C$ | $f_{IU}^C$ | $k_{FI}^C$ | $k_{IU}^C$ | $x_{FI}^{\ddagger}$ | $x_{IF}^{\ddagger}$ | $x_{IU}^{\ddagger}$ | $x_{UI}^{\ddagger}$ |
|---|---|---|---|---|---|---|---|---|
| Mol1 | 14.60 | 12.88 | 6.59 | 9.18 | 7.78 | 7.33 | 6.16 | 3.26 |
| Mol2 | 14.63 | 12.91 | 8.48 | 9.80 | 7.44 | 5.98 | 6.03 | 4.22 |
| Mol3 | 14.60 | 12.88 | 6.66 | 9.29 | 7.83 | 7.25 | 6.11 | 3.36 |
| Mol4 | 14.83 | 12.98 | 7.68 | 9.29 | 7.67 | 6.98 | 5.92 | 3.67 |
| Mol5 | 14.82 | 12.91 | 5.54 | 7.37 | 6.19 | 6.40 | 6.40 | 5.90 |
| Mol6 | 14.00 | 12.31 | 7.53 | 8.48 | 7.69 | 6.61 | 6.06 | 4.72 |
| Mol7 | 14.73 | 12.69 | 4.50 | 8.49 | 7.36 | 6.38 | 5.01 | 5.49 |
| **Average** | **14.6** | **12.8** | **6.71** | **8.84** | **7.42** | **6.71** | **5.95** | **4.4** |
| Statistical error | 0.11 | 0.09 | 0.51 | 0.30 | 0.22 | 0.19 | 0.17 | 0.39 |
|  | $x_{FI}$ | $x_{IU}$ | $x_{FU}$ | $\Delta G_{FI}$ | $\Delta G_{IU}$ | | | |
| Mol1 | 15.11 | 9.42 | 24.54 | 53.68 | 29.53 | | | |
| Mol2 | 13.42 | 10.25 | 23.67 | 47.78 | 32.18 | | | |
| Mol3 | 15.08 | 9.47 | 24.55 | 53.58 | 29.67 | | | |
| Mol4 | 14.66 | 9.59 | 24.25 | 52.90 | 30.30 | | | |
| Mol5 | 12.59 | 12.30 | 24.88 | 45.40 | 38.63 | | | |
| Mol6 | 14.31 | 10.78 | 25.09 | 48.75 | 32.29 | | | |
| Mol7 | 13.75 | 10.49 | 24.24 | 49.28 | 32.41 | | | |
| **Average** | **14.1** | **10.3** | **24.5** | **50.2** | **32.1** | | | |
| Statistical error | 0.35 | 0.38 | 0.18 | 1.22 | 1.18 | | | |

| 3S SH CFM Bell-Evans | $f_{FI}^C$ | $f_{IU}^C$ | $k_{FI}^C$ | $k_{IU}^C$ | $x_{FI}^{\ddagger}$ | $x_{IF}^{\ddagger}$ | $x_{IU}^{\ddagger}$ | $x_{UI}^{\ddagger}$ |
|---|---|---|---|---|---|---|---|---|
| Mol1 | 14.49 | 13.31 | 5.93 | 5.34 | 5.00 | 10.46 | 11.22 | 3.70 |
| Mol2 | 14.72 | 13.29 | 7.39 | 6.63 | 6.96 | 8.65 | 8.74 | 5.78 |
| Mol3 | 14.04 | 12.97 | 4.33 | 6.28 | 7.80 | 12.90 | 11.18 | 2.38 |
| Mol4 | 14.22 | 12.18 | 8.81 | 5.97 | 7.10 | 3.65 | 5.92 | 3.88 |
| Mol5 | 13.75 | 13.08 | 3.92 | 18.02 | 5.26 | 17.53 | 9.68 | 9.30 |
| Mol6 | 14.56 | 13.11 | 9.52 | 5.34 | 7.55 | 6.64 | 10.61 | 4.53 |
| Mol7 | 14.55 | 13.00 | 6.86 | 5.54 | 5.43 | 7.78 | 7.86 | 4.34 |
| **Average** | **14.33** | **12.99** | **6.68** | **7.59** | **6.44** | **9.66** | **9.32** | **4.85** |
| Statistical error | 0.13 | 0.14 | 0.80 | 1.75 | 0.44 | 1.71 | 0.74 | 0.84 |
|  | $x_{FI}$ | $x_{IU}$ | $x_{FU}$ | $\Delta G_{FI}$ | $\Delta G_{IU}$ | | | |
| Mol1 | 15.47 | 14.92 | 30.38 | 54.53 | 48.30 | | | |
| Mol2 | 15.61 | 14.53 | 30.13 | 55.89 | 46.99 | | | |
| Mol3 | 20.70 | 13.57 | 34.27 | 70.75 | 42.81 | | | |
| Mol4 | 10.76 | 9.80 | 20.55 | 37.21 | 29.05 | | | |
| Mol5 | 22.79 | 18.98 | 41.77 | 76.22 | 60.40 | | | |
| Mol6 | 14.18 | 15.14 | 29.32 | 50.26 | 48.29 | | | |
| Mol7 | 13.21 | 12.20 | 25.41 | 46.78 | 38.58 | | | |
| **Average** | **16.10** | **14.16** | **30.26** | **55.95** | **44.92** | | | |
| Statistical error | 1.60 | 1.07 | 2.52 | 5.12 | 3.66 | | | |

| 3S SH PM Detailed Balance | $f^C_{FI}$ | $f^C_{IU}$ | $x_{FI}$ | $x_{IU}$ | $x_{FU}$ | $\Delta G_{FI}$ | $\Delta G_{IU}$ |
|---|---|---|---|---|---|---|---|
| Mol1 | 14.55 | 12.79 | 15.54 | 8.95 | 24.50 | 55.03 | 27.85 |
| Mol2 | 14.56 | 12.86 | 13.75 | 9.98 | 23.73 | 48.70 | 31.23 |
| Mol3 | 14.55 | 12.79 | 15.47 | 9.02 | 24.49 | 54.77 | 28.07 |
| Mol4 | 14.78 | 12.97 | 14.61 | 9.79 | 24.40 | 52.55 | 30.89 |
| Mol5 | 14.65 | 13.67 | 16.50 | 10.40 | 26.90 | 58.82 | 34.59 |
| Mol6 | 13.93 | 12.30 | 14.67 | 10.49 | 25.17 | 49.74 | 31.40 |
| Mol7 | 14.65 | 12.74 | 14.21 | 10.19 | 24.41 | 50.65 | 31.61 |
| **Average** | **14.5** | **12.9** | **15.0** | **9.83** | **24.8** | **52.9** | **30.8** |
| Statistical error | 0.10 | 0.15 | 0.35 | 0.24 | 0.38 | 1.35 | 0.87 |

| 3S SH CFM Detailed Balance | $f^C_{FI}$ | $f^C_{IU}$ | $x_{FI}$ | $x_{IU}$ | $x_{FU}$ | $\Delta G_{FI}$ | $\Delta G_{IU}$ |
|---|---|---|---|---|---|---|---|
| Mol1 | 14.48 | 13.30 | 15.72 | 14.68 | 30.40 | 55.39 | 47.51 |
| Mol2 | 14.73 | 13.29 | 15.56 | 14.60 | 30.15 | 55.76 | 47.20 |
| Mol3 | 14.16 | 13.03 | 16.11 | 15.40 | 31.50 | 55.49 | 48.83 |
| Mol4 | 14.20 | 12.17 | 10.91 | 9.72 | 20.63 | 37.70 | 28.78 |
| Mol5 | 13.86 | 12.85 | 18.83 | 13.75 | 32.58 | 63.47 | 42.99 |
| Mol6 | 14.57 | 13.09 | 14.25 | 14.87 | 29.12 | 50.54 | 47.34 |
| Mol7 | 14.56 | 12.99 | 13.27 | 12.14 | 25.41 | 47.01 | 38.36 |
| **Average** | **14.37** | **12.96** | **14.95** | **13.59** | **28.54** | **52.19** | **43.00** |
| Statistical error | 0.12 | 0.14 | 0.94 | 0.76 | 1.57 | 3.09 | 2.74 |

*Table S9d. Kinetic and thermodynamic parameters for the 3S hairpin with long handles.*

| 3S LH PM Bell-Evans | $f^C_{FI}$ | $f^C_{IU}$ | $k^C_{FI}$ | $k^C_{IU}$ | $x^‡_{FI}$ | $x^‡_{IF}$ | $x^‡_{IU}$ | $x^‡_{UI}$ |
|---|---|---|---|---|---|---|---|---|
| Mol1 | 14.35 | 12.27 | 15.32 | 12.49 | 8.22 | 5.39 | 4.99 | 4.59 |
| Mol2 | 14.91 | 12.90 | 18.82 | 12.86 | 8.15 | 5.09 | 5.45 | 4.11 |
| Mol3 | 14.90 | 12.78 | 21.82 | 11.39 | 8.70 | 3.81 | 6.20 | 3.75 |
| Mol4 | 13.98 | 12.45 | 11.51 | 9.70 | 8.65 | 4.99 | 5.49 | 4.31 |
| Mol5 | 14.37 | 12.38 | 22.92 | 12.45 | 7.61 | 4.52 | 5.88 | 3.49 |
| **Average** | **14.50** | **12.56** | **18.08** | **11.78** | **8.27** | **4.76** | **5.60** | **4.05** |
| Statistical error | 0.178 | 0.12 | 2.11 | 0.57 | 0.19 | 0.27 | 0.20 | 0.19 |

| | $x_{FI}$ | $x_{IU}$ | $x_{FU}$ | $\Delta G_{FI}$ | $\Delta G_{IU}$ |
|---|---|---|---|---|---|
| Mol1 | 13.62 | 9.58 | 23.21 | 47.56 | 28.62 |
| Mol2 | 13.24 | 9.57 | 22.81 | 48.04 | 30.05 |
| Mol3 | 12.51 | 9.96 | 22.48 | 45.38 | 30.97 |
| Mol4 | 13.64 | 9.80 | 23.44 | 46.42 | 29.70 |
| Mol5 | 12.14 | 9.37 | 21.51 | 42.46 | 28.22 |
| **Average** | **13.03** | **9.66** | **22.69** | **45.97** | **29.51** |
| Statistical error | 0.30 | 0.10 | 0.33 | 0.99 | 0.49 |

| 3S LH CFM Bell-Evans | $f^C_{FI}$ | $f^C_{IU}$ | $k^C_{FI}$ | $k^C_{IU}$ | $x^{\ddagger}_{FI}$ | $x^{\ddagger}_{IF}$ | $x^{\ddagger}_{IU}$ | $x^{\ddagger}_{UI}$ |
|---|---|---|---|---|---|---|---|---|
| Mol1 | 14.97 | 13.32 | 16.13 | 10.98 | 9.29 | 4.94 | 6.11 | 7.45 |
| Mol2 | 14.57 | 13.15 | 12.19 | 8.50 | 10.33 | 8.97 | 6.75 | 6.04 |
| Mol3 | 14.57 | 12.85 | 9.54 | 8.70 | 6.22 | 6.31 | 6.70 | 7.49 |
| Mol4 | 13.94 | 12.52 | 15.86 | 10.36 | 10.76 | 6.73 | 8.00 | 8.55 |
| Mol5 | 14.11 | 12.74 | 19.36 | 7.69 | 9.79 | 5.27 | 9.48 | 6.66 |
| **Average** | **14.43** | **12.92** | **14.62** | **9.25** | **9.28** | **6.44** | **7.41** | **7.24** |
| Statistical error | 0.189 | 0.14 | 1.70 | 0.61 | 0.80 | 0.71 | 0.60 | 0.42 |
|  | $x_{FI}$ | $x_{IU}$ | $x_{FU}$ | $\Delta G_{FI}$ | $\Delta G_{IU}$ |  |  |  |
| Mol1 | 14.23 | 13.57 | 27.81 | 51.86 | 44.00 |  |  |  |
| Mol2 | 19.31 | 12.79 | 32.10 | 68.46 | 40.96 |  |  |  |
| Mol3 | 12.53 | 14.20 | 26.73 | 44.44 | 44.42 |  |  |  |
| Mol4 | 17.49 | 16.56 | 34.05 | 59.35 | 50.47 |  |  |  |
| Mol5 | 15.07 | 16.14 | 31.21 | 51.75 | 50.07 |  |  |  |
| **Average** | **15.72** | **14.65** | **30.38** | **55.17** | **45.98** |  |  |  |
| Statistical error | 0.30 | 0.73 | 1.36 | 4.07 | 1.85 |  |  |  |

| 3S LH PM Detailed Balance | $f^C_{FI}$ | $f^C_{IU}$ | $x_{FI}$ | $x_{IU}$ | $x_{FU}$ | $\Delta G_{FI}$ | $\Delta G_{IU}$ |
|---|---|---|---|---|---|---|---|
| Mol1 | 14.23 | 12.28 | 13.87 | 9.33 | 23.21 | 47.96 | 27.93 |
| Mol2 | 14.82 | 12.84 | 12.79 | 9.15 | 21.95 | 46.10 | 28.65 |
| Mol3 | 14.85 | 12.64 | 11.21 | 9.21 | 20.43 | 40.50 | 28.41 |
| Mol4 | 14.09 | 12.14 | 13.20 | 9.70 | 22.90 | 45.24 | 28.67 |
| Mol5 | 14.20 | 12.26 | 13.00 | 8.64 | 21.64 | 44.88 | 25.82 |
| **Average** | **14.44** | **12.43** | **12.82** | **9.21** | **22.03** | **44.94** | **27.90** |
| Statistical error | 0.16 | 0.13 | 0.43 | 0.17 | 0.49 | 1.23 | 0.53 |

| 3S LH CFM Detailed Balance | $f^C_{FI}$ | $f^C_{IU}$ | $x_{FI}$ | $x_{IU}$ | $x_{FU}$ | $\Delta G_{FI}$ | $\Delta G_{IU}$ |
|---|---|---|---|---|---|---|---|
| Mol1 | 14.95 | 13.32 | 14.29 | 13.49 | 27.79 | 52.02 | 43.78 |
| Mol2 | 14.60 | 13.15 | 18.10 | 12.46 | 30.57 | 64.34 | 39.89 |
| Mol3 | 14.54 | 12.56 | 13.12 | 11.30 | 24.43 | 46.43 | 34.57 |
| Mol4 | 13.92 | 12.52 | 17.66 | 16.31 | 33.98 | 59.86 | 49.73 |
| Mol5 | 14.16 | 12.75 | 14.09 | 16.07 | 30.17 | 48.60 | 49.89 |
| **Average** | **14.44** | **12.86** | **15.45** | **13.93** | **29.39** | **54.25** | **43.57** |
| Statistical error | 0.17 | 0.16 | 1.01 | 0.98 | 1.58 | 3.40 | 2.93 |

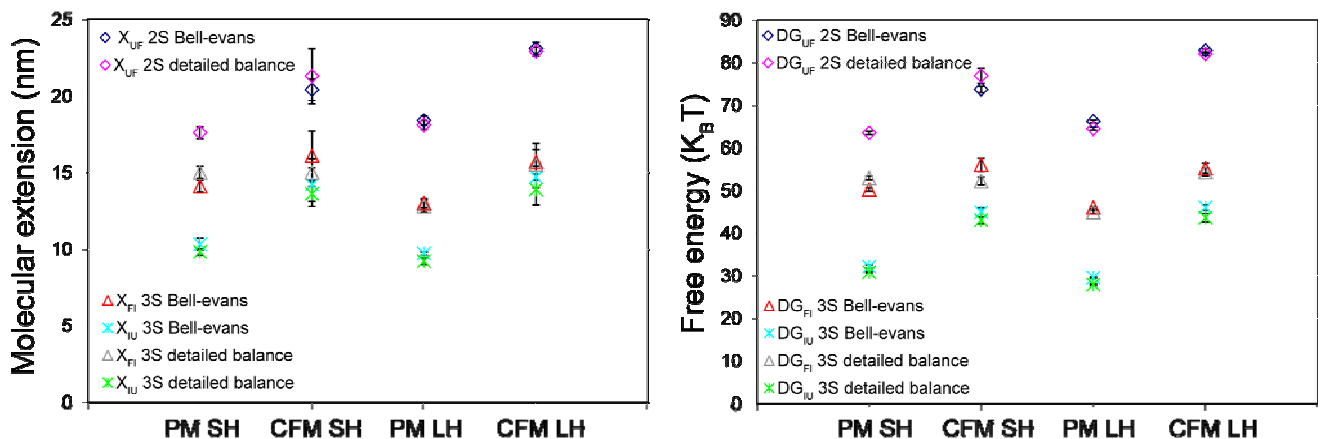

FIGURE S9: Results for the molecular distances (left panel) and free energies (right panel) as estimated from the Bell-Evans and detailed balance condition for the different experimental conditions tested (PM SH, CFM SH, PM LH, CFM LH). Results correspond to the average over 4-7 molecules (Table 1 main text).